\documentclass[12pt]{article}
\usepackage{amsmath}
\usepackage{epsfig}
\usepackage{psfrag}

\def\no{\nonumber}
\def\eps{\varepsilon}
\def\Li{\text{Li}}
\def\S{\text{S}}
\def\as{\alpha_s}
\def\LQCD{\Lambda_\text{QCD}}
\def\MSbar{\ensuremath{\overline{\text{MS}}}}
\def\calI{\mathcal{I}}
\def\calP{\mathcal{P}}
\def\calO{\mathcal{O}}
\def\calC{\mathcal{C}}
\def\calH{\mathcal{H}}
\def\calQ{\mathcal{Q}}
\def\ub{\bar{u}}

\allowdisplaybreaks[1]

\begin{document}


\begin{titlepage}

\begin{flushright}
 TTP08-49\\
 SFB/CPP-08-90
\end{flushright}
\vskip 2.2cm

\begin{center}
\Large\bf\boldmath
NNLO corrections to inclusive semileptonic\\
$B$ decays in the shape-function region \unboldmath

\normalsize
\vskip 1.5cm

{\sc Guido~Bell\footnote{E-mail:bell@particle.uni-karlsruhe.de}}

\vskip .5cm

{\it Institut f\"ur Theoretische Teilchenphysik, \\
Universit\"at Karlsruhe, D-76128 Karlsruhe, Germany}

\vskip 1.5cm

\end{center}

\begin{abstract}
\noindent
We compute 2-loop QCD corrections to the hard coefficient functions
which arise in the factorization formula for $B\to X_u\ell\nu$ decays in
the shape-function region. Our calculation provides the last missing
piece required for a NNLO analysis of inclusive semileptonic $B$ decays,
which may significantly reduce the theoretical uncertainty in the
extraction of the CKM matrix element $|V_{ub}|$. Among the technical
aspects, we find that the 2-loop hard coefficient functions are free of
infrared singularities as predicted by the factorization framework. We
perform a brief numerical analysis of the NNLO corrections and include a
discussion on charm mass effects.   
\end{abstract}

\vfill

\end{titlepage}


\section{Introduction}


Charmless semileptonic $B$ meson decays are mediated by a $b\to u$ quark
transition. The study of inclusive and exclusive semileptonic $B$ decays
provide two independent paths for a determination of the strength of the
flavour-changing interaction. In the Standard Model the quark transition
is governed by the CKM matrix element $V_{ub}$, which constitutes an
important input parameter for many analyses in flavour physics. 

Since the determination of CKM matrix elements from exclusive
semileptonic $B$ meson decays requires the knowledge of non-perturbative
form factors, the determination from inclusive semileptonic decays is a
priori cleaner from the theoretical point of view. The total decay rate
can be described by a local Operator Product Expansion (OPE) in a
simultaneous expansion in $\LQCD/m_b$ and
$\alpha_s(m_b)$~\cite{Chay:1990da}. The local OPE can be applied for
inclusive $B\to X_c\ell\nu$ decays, where the non-perturbative
information is constrained to some numbers, HQET parameters, which can
be extracted simultaneously with $|V_{cb}|$ from an analysis of the $B$
meson decay spectra. 

A major complication arises in charmless semileptonic $B$ decays where
experimental measurements of the $B\to X_u\ell\nu$ decay rate have to
introduce kinematical cuts to suppress the $B\to X_c\ell\nu$
background. This restricts the experimentally accessible information to
the shape-function region in which the hadronic final state has large
energy $E_X \sim m_b$ but moderate invariant mass $p_X^2 \sim m_b
\LQCD$. The theoretical description of partial decay rates in the
shape-function region is more complicated and gives rise to a
multi-scale OPE in terms of non-local light-cone operators. 

The development of QCD Factorization\cite{BBNS} and Soft-Collinear
Effective Theory(SCET) \cite{SCET} has widely improved our understanding
of strong interaction effects in charmless $B$ meson decays. A
factorization theorem for the partial $B\to X_u\ell\nu$ decay rate in
the shape-function region has originally been proposed in the
diagrammatic approach~\cite{Korchemsky:1994jb} and has later been
formulated in the operator formalism within SCET. The factorization
formula can be expressed in terms of the hadronic tensor $W^{\mu\nu}$,
which encodes the strong interaction effects in $B\to X_u\ell\nu$
decays, schematically as   
\begin{align}
W^{\mu\nu} \sim \sum_{i,j} \; c_{i j}^{\mu \nu} \, H_{i j}(n_+ p) \,
\int d\omega \; J(p_\omega^2) \, S(\omega),
\label{eq:ff:short}
\end{align}
with hard coefficient functions $H_{i j}$, a jet function $J$ and a
shape function $S$ which describes the distribution of the residual
light-cone momentum of the $b$-quark within the $B$
meson~\cite{ShapeFunc} (the $c's$ are some tensor coefficients). Whereas
the hard functions and the jet function can be computed in perturbation
theory, since they describe fluctuations with virtualities $\sim m_b^2$
and $\sim \LQCD m_b$ respectively, the shape function represents a
non-perturbative input to the factorization formula. 

In this work we address perturbative corrections to the factorization
formula (\ref{eq:ff:short}). Next-to-leading order (NLO) corrections to
the hard functions $H_{ij}$ and the jet function $J$ have been computed
in~\cite{NLO} and next-to-next-to-leading order (NNLO) corrections to
the jet function have been worked out in~\cite{Becher:2006qw}. The only
missing piece for a NNLO analysis of inclusive semileptonic $B$ decays
in the shape-function region consists in the $\alpha_s^2$ corrections to
the hard functions which we consider in this work. 

A first step towards the computation of NNLO corrections to the hard
functions has been taken recently in~\cite{Bonciani:2008wf}. The authors
of~\cite{Bonciani:2008wf} computed 2-loop QCD corrections to the form
factors which parameterize the $b\to u$ quark transition. We present an
independent calculation of these 2-loop form factors and extend the
analysis of~\cite{Bonciani:2008wf} in two respects: First, we perform
the subsequent matching calculation within SCET to extract the hard
functions $H_{ij}$ from the form factors which are formally infrared
divergent. Second, we keep a non-vanishing charm quark mass in our
2-loop calculation which has been set to zero
in~\cite{Bonciani:2008wf}. 

The 2-loop calculation that we present in this work is closely related
to the calculation of NNLO vertex corrections in hadronic $B$ decays
which we considered in~\cite{GB}. It is in fact somewhat simpler than
the one in~\cite{GB} since it involves only a small subset of Feynman
diagrams and 2-loop integrals. A further complication arises in the
analysis of~\cite{GB} through the appearance of evanescent four-quark
operators which makes the matching calculation rather involved.

The outline of this paper is as follows: To set up our notation, we
briefly recapitulate the necessary ingredients for the matching
calculation in Section~\ref{sec:setup}. The 2-loop calculation of the
QCD form factors is described in Section~\ref{sec:QCD}. We perform the
matching calculation in Section~\ref{sec:matching}, where we present our
results for the NNLO hard coefficient functions $H_{ij}$. We extend our
analysis to include charm mass effects in Section~\ref{sec:charm}. We
briefly analyze the numerical impact of the NNLO corrections in
Section~\ref{sec:numerics}, before we conclude in
Section~\ref{sec:concl}. We collect some results for the NLO and NNLO
coefficient functions and for the charm mass dependent Master Integrals
in Appendices~\ref{app:1loop}-\ref{app:MIs}.


\section{Preliminaries}

\label{sec:setup}

The basic quantity for the analysis of strong interaction effects to
inclusive semileptonic $B$ meson decays is the hadronic tensor
$W^{\mu\nu}$, which is defined via the discontinuity of the forward
matrix element of the correlation function of two weak interaction
currents $J^\mu=\bar{u} \gamma^\mu(1-\gamma_5)b$, 
\begin{align}
W^{\mu\nu} = \frac{1}{\pi} \; \text{Im} \,
\langle B(p_B) | \, i \int d^4x \;e^{i q x} \;
\text{T} \left[ J^{\dagger\mu}(0) J^{\nu}(x) \right] | B(p_B) \rangle.
\end{align}
For the present analysis of inclusive semileptonic decays in the
shape-function region it will be convenient to introduce light-cone
coordinates. We introduce two light-like vectors $n_\pm$ with
$n_\pm^2=0$ and $n_+ n_-=2$. Any four-vector can be decomposed according
to its projections onto these light-cone directions and a
two-dimensional transverse plane as 
\begin{align}
p^\mu = (n_+ p) \frac{n_-^\mu}{2} + p_\perp^\mu +
(n_- p) \frac{n_+^\mu}{2} \equiv
(n_+p,p_\perp,n_-p).
\end{align}
In the shape-function region the momentum of the hadronic final state
$X_u$ scales as a hard-collinear momentum, i.e.~$p_X\sim
m_b(1,\lambda,\lambda^2)$ with $\lambda^2=\LQCD/m_b$. A factorization
theorem for the hadronic tensor $W^{\mu\nu}$ can be established in a
two-step matching procedure QCD~$\to$~SCET~$\to$~HQET. In the first step
hard fluctuations with virtuality $\sim m_b^2$ are integrated out. The
semileptonic current becomes to leading power in $\lambda$
\begin{align}
J^\mu (x) \simeq
e^{-i m_b v\cdot x} \sum_{i=1}^{3} \int ds \; \tilde{C}_i(s)
\left[\bar{\xi}W_{hc}\right](x+sn_+)\Gamma_i^\mu
h_v(x_-),
\label{eq:matching}
\end{align}
with the $B$ meson velocity $v$ and $x_-=(n_+x)n_-/2$. The static
$b$-quark field $h_v$ is defined in HQET and the hard-collinear
$u$-quark field $\xi$ and the Wilson line $W_{hc}$ in SCET. The momentum
space Wilson coefficients can be obtained via
\begin{align}
C_i(n_+p) = \int ds \; e^{is n_+ p} \, \tilde{C}_i(s).
\end{align}
In the second matching step hard-collinear fluctuations with virtuality
$\sim \LQCD m_b$ are integrated out. This gives rise to another
perturbative coefficient function, the jet function $J$, and a remnant
HQET matrix element, the shape function $S$. The factorization formula
for the hadronic tensor becomes to leading power in $\lambda$ 
\begin{align}
W^{\mu\nu} \simeq \sum_{i,j=1}^{3} \; H_{i j}(n_+ p) \;
\text{Tr}\left(\bar{\Gamma}_j^\mu \frac{p\!\!\!\slash}{2} \Gamma_i^\nu
  \frac{1+v\!\!\!\slash}{2} \right) \,
\int d\omega \; J(p_\omega^2) \, S(\omega).
\label{eq:ff:exact}
\end{align}
The hard functions $H_{ij}$, which we consider in this work, are related
to the Wilson coefficients $C_i$ of the semileptonic current by
$H_{ij}=C_i C_j$. We may therefore mainly focus on the matching relation
(\ref{eq:matching}) instead of the factorization formula for the
hadronic tensor (\ref{eq:ff:exact}). 

The matching calculation simplifies if we work with on-shell quarks and
regularize ultraviolet (UV) and infrared (IR) singularities in
Dimensional Regularization (DR) (we write $d=4-2\eps$ and use an
anticommuting $\gamma_5$ according to the NDR scheme). In this case the
SCET diagrams are scaleless and vanish and the main task consists in the
computation of the 2-loop QCD corrections to the left-hand side of
(\ref{eq:matching}). We parameterize the QCD result in terms of three
form factors, 
\begin{align}
\langle u(p) | J^\mu | b(p_b) \rangle
= \bar{u}(p) \left[
F_1(q^2) \gamma^\mu(1-\gamma_5)
+ F_2(q^2) \frac{p_b^\mu}{m_b} (1+\gamma_5)
+ F_3(q^2) \frac{p^\mu}{m_b} (1+\gamma_5)
\right] b(p_b),
\end{align}
with momentum transfer $q=p_b-p$. On the other hand we work in a frame
with $v=(n_-+n_+)/2$ and choose the basis of three independent Dirac
structures in (\ref{eq:matching}) as 
\begin{align}
\Gamma_1^\mu=\gamma^\mu(1-\gamma_5),\qquad
\Gamma_2^\mu=v^\mu(1+\gamma_5),\qquad
\Gamma_3^\mu=n_-^\mu(1+\gamma_5).
\end{align}
Writing the momentum of the on-shell $b$-quark as $p_b=m_b v$ and the
one of the $u$-quark as $p = (n_+p)n_-/2$ provides the link between the
form factors $F_i$ and the coefficient functions $C_i$. We finally
introduce a dimensionless variable $u$ to parameterize the light-cone
momentum of the $u$-quark as $u=n_+p/m_b$. In this notation the momentum
transfer becomes $q^2=\ub m_b^2$ with $\ub=1-u$.


\section{NNLO calculation of QCD form factors}

\label{sec:QCD}

\subsection{2-loop calculation}
\label{sec:2loop}

\begin{figure}[b!]
\centerline{\parbox{14cm}{\centerline{
\includegraphics[width=10cm]{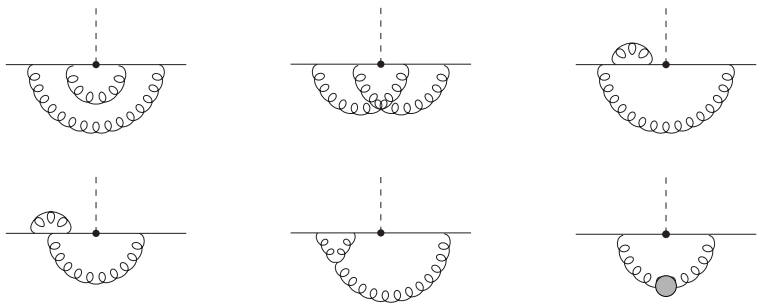}}
\caption{\label{fig:Diagrams} \small \textit{2-loop diagrams. Diagrams
    that result from mirroring at the vertical axis are not shown. The
    bubble in the last diagram represents the 1-loop gluon
    self-energy.}}}} 
\end{figure}

To NNLO the QCD calculation of the form factors $F_i(q^2)$ involves the
calculation of the 2-loop diagrams from Figure~\ref{fig:Diagrams}. As
pointed out in the introduction, the calculation can be inferred from
the calculation of the 2-loop vertex corrections to hadronic $B$
decays~\cite{GB}. The present calculation involves only a small subset
of 2-loop diagrams and about half of the Master Integrals (MIs) that
have been computed in~\cite{GB}. 

\begin{figure}[t!]
\centerline{\parbox{14cm}{\centerline{
\includegraphics[width=14cm]{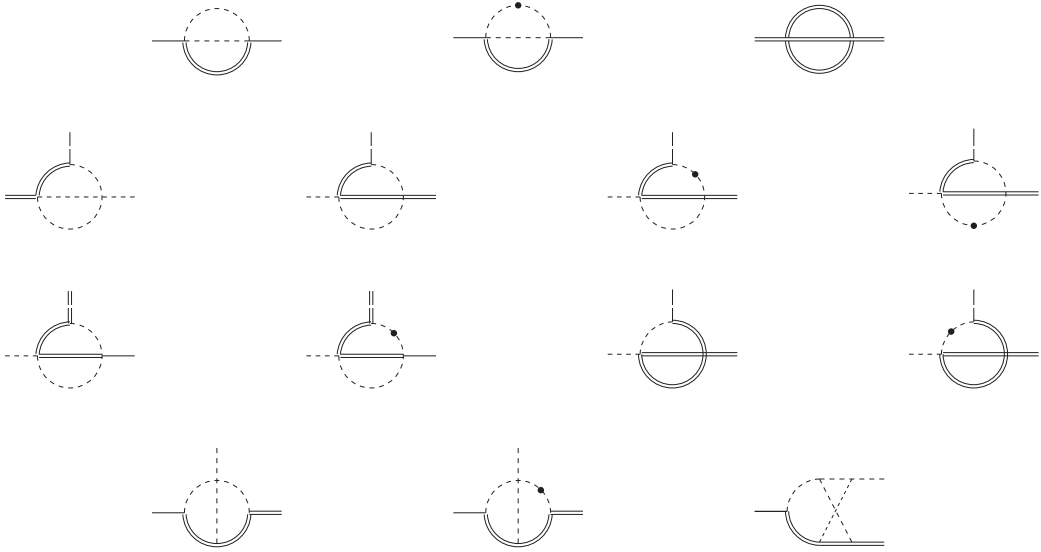}}
\caption{\label{fig:MIs} \small \textit{Scalar 2-loop Master
    Integrals. We use dashed (double) lines for massless (massive)
    propagators. Dashed/solid/double external lines correspond to
    virtualities $0/u m_b^2/m_b^2$, respectively. Dotted propagators are
    taken to be squared.}}}} 
\end{figure}

The calculation follows the same strategy that has been used
in~\cite{GB}. We first express tensor integrals in terms of scalar
integrals with the help of a general tensor decomposition. We then use
an automatized reduction algorithm, which is based on
integration-by-parts techniques~\cite{IBP} and the Laporta
algorithm~\cite{Laporta}, to express the scalar integrals in terms of an
irreducible  set of Master Integrals (MIs). As long as we neglect the
charm quark mass, the present calculation gives rise to 14 MIs which are
depicted in Figure~\ref{fig:MIs}. All MIs have already been calculated
in~\cite{GB} with the help of the method of differential
equations~\cite{DiffEqs}, the formalism of Harmonic Polylogarithms
(HPLs)~\cite{HPLs}, Mellin-Barnes techniques~\cite{MB} and the method of
sector decomposition~\cite{SecDecomp}. 

The MIs from Figure~\ref{fig:MIs} can be expressed in terms of HPLs of
weight $w\leq4$. Throughout this paper, we give our results in terms of
the following minimal set of HPLs 
\begin{align}
H(0;x) &= \ln(x),  &
H(0,0,1;x) &= \Li_{3}(x),
\no \\
H(1;x) &= -\ln(1-x), &
H(0,1,1;x) &= \S_{1,2}(x),
\no \\
H(-1;x) &= \ln(1+x), &
H(0,0,0,1;x) &= \Li_4(x),
\no \\
H(0,1;x) &= \Li_2(x), &
H(0,0,1,1;x) &= \S_{2,2}(x),
\no \\
H(0,-1;x) &= -\Li_2(-x), &
H(0,1,1,1;x) &= \S_{1,3}(x).
\end{align}
Moreover, we introduce a shorthand notation for a HPL of weight $w=3$,
whose expression in terms of Nielsen Polylogarithms is rather involved 
\begin{align}
\calH_1(x) &\equiv H(-1,0,1;x ) \no\\
&=
\Li_3 \Big(\frac{1-x}{2}\Big) + \Li_3 \Big(\frac{1-x}{1+x}\Big)
+ \S_{1,2}(x) + \S_{1,2}(-x) - \S_{1,2} \Big(\frac{1-x}{2}\Big)
\no\\
&\quad + \ln(1-x) \Li_2(x) - \ln \Big(\frac{1-x}{1+x}\Big) \Li_2(-x)
- \ln \Big(\frac{1+x}{2}\Big) \Li_2 \Big(\frac{1-x}{2}\Big)
\no\\
&\quad - \frac16 \ln^3 \Big(\frac{1+x}{2}\Big)
+ \frac12 \ln(x) \ln^2 \Big(\frac{1-x}{1+x}\Big)
- \frac{\pi^2}{4} \ln \Big(\frac{1-x}{1+x}\Big) - \frac74 \zeta_3,
\end{align}
and for a HPL of weight $w=4$, which cannot be expressed in terms of
Nielsen Poly-logarithms and has to be evaluated numerically, 
\begin{align}
\calH_2(x) \equiv H(0,-1,0,1;x ) = \int_0^x dx' \,
\frac{\calH_1(x')}{x'}.
\end{align}
The MIs from Figure~\ref{fig:MIs} have been computed in an analytic
form~\cite{GB}, apart from a single number $\calC_0$ which arises in the
calculation of the boundary condition to the 6-topology MI (last diagram
from Figure~\ref{fig:MIs}). With the help of the {\sc Mathematica}
package {\sc FIESTA}~\cite{FIESTA}, we can compute this number to higher
precision than in~\cite{GB}. We now find $\calC_0 = -60.250 \pm 0.001$.

\subsection{Renormalization}

\begin{figure}[b!]
\centerline{\parbox{14cm}{
\centerline{\includegraphics[width=4.2cm]{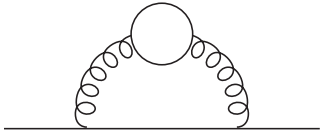}}
\caption{\label{fig:Z2u}  \small \textit{2-loop diagram with massive
    fermion loop.}}}} 
\end{figure}

The form factors, which we obtain from the 2-loop diagrams of
Figure~\ref{fig:Diagrams}, are UV- and IR-divergent. The UV-divergences
are subtracted in the renormalization procedure which amounts to the
calculation of several 1-loop diagrams with standard QCD counterterm
insertions. We account for the wave-function renormalization of the
quark fields by 
\begin{align}
F_{i} &= Z_{2,u}^{1/2} Z_{2,b}^{1/2} \, F_{i,\text{bare}},
\end{align}
where $Z_{2,u}$ ($Z_{2,b}$) is the wave-function renormalization factor
of the massless $u$-quark (massive $b$-quark). Writing the perturbative
expansion in terms of the renormalized coupling constant $\as$, 
\begin{align}
F_{i,(\text{bare})} =
\sum_{k=0}^\infty \left( \frac{\as}{4\pi} \right)^k
F_{i,(\text{bare})}^{(k)},
\hspace{1.5cm}
Z_{2,u/b} = 1 + \sum_{k=1}^\infty  \left( \frac{\as}{4\pi} \right)^k
Z_{2,u/b}^{(k)},
\label{eq:pertexp}
\end{align}
we find that the renormalized form factors are given to NNLO by
\begin{align}
F_{i}^{(0)} &= F_{i,\text{bare}}^{(0)}, \no \\
F_{i}^{(1)} &= F_{i,\text{bare}}^{(1)}
  + \frac12 \bigg( Z_{2,u}^{(1)} + Z_{2,b}^{(1)} \bigg)
    F_{i,\text{bare}}^{(0)}, \no \\
F_{i}^{(2)} &= F_{i,\text{bare}}^{(2)}
  + \frac12 \bigg( Z_{2,u}^{(1)} + Z_{2,b}^{(1)} \bigg)
    F_{i,\text{bare}}^{(1)}
  + \bigg( \frac12 \big( Z_{2,u}^{(2)} + Z_{2,b}^{(2)} \big)
    - \frac18 \big( Z_{2,u}^{(1)} - Z_{2,b}^{(1)} \big)^2 \bigg)
    F_{i,\text{bare}}^{(0)}.
\label{eq:expFren}
\end{align}
We renormalize the coupling constant in the \MSbar-scheme (adopting the
notation (\ref{eq:pertexp})) 
\begin{align}
Z_g^{(1)} = - \left( \frac{11}{6}C_A - \frac23 n_f T_F \right)
\frac{1}{\eps},
\end{align}
whereas the quark wave-functions and the $b$-quark mass are renormalized
in the on-shell scheme, 
\begin{align}
Z_{2,b}^{(1)} = Z_{m}^{(1)} = - C_F
\left( \frac{e^{\gamma_E}\mu^2}{m_b^2} \right)^\eps \Gamma(\eps) \;
\frac{3-2\eps}{1-2\eps},
\end{align}
and $Z_{2,u}^{(1)}=0$ because of the fact that scaleless integrals
vanish in DR. According to (\ref{eq:expFren}), we also need the 2-loop
expressions for the wave-function renormalization factors. For a
massless quark $Z_{2,u}^{(2)}$ can be calculated easily, since it
receives a contribution from a single diagram, shown in
Figure~\ref{fig:Z2u}, which introduces a mass scale. We find, in
agreement with e.g.~\cite{Beneke:2008ei}, 
\begin{align}
Z_{2,u}^{(2)} = T_F C_F
\left( \frac{e^{\gamma_E}\mu^2}{m_b^2} \right)^{2\eps} \Gamma(\eps)^2 \;
\frac{2\eps(3-2\eps)(1+\eps)}{(1-\eps)(2-\eps)(1+2\eps)(3+2\eps)}.
\end{align}
For a massive quark $Z_{2,b}^{(2)}$ has been calculated
in~\cite{Z2bOS:2loop}, 
\begin{align}
Z_{2,b}^{(2)} &= C_F
\left\{
\left[ \frac92 C_F + \frac{11}{2} C_A - 2 n_l T_F  \right]
\frac{1}{\eps^2}  \right.\no\\
& \quad
+ \left[ \bigg( \frac{51}{4} + 9 L \bigg) C_F - \frac{127}{12} C_A +
  \frac{11}{3} n_l T_F + \bigg(1 + 4 L\bigg)  T_F \right]
\frac{1}{\eps}  \no\\
& \quad
+ \left( \frac{433}{8} - \frac{49\pi^2}{4} +  16 \pi^2 \ln(2) - 24
  \zeta_3 + \frac{51}{2} L + 9 L^2 \right) C_F \no\\
& \quad
+ \left( -\frac{1705}{24} + 5\pi^2 - 8 \pi^2 \ln(2) + 12 \zeta_3 -
  \frac{215}{6} L - \frac{11}{2} L^2\right) C_A \no\\
& \quad \left.
+  \left( \frac{113}{6} + \frac{4\pi^2}{3} +\frac{38}{3} L +2 L^2
\right) n_l T_F
+ \left( \frac{947}{18} - 5 \pi^2 + \frac{22}{3} L + 6 L^2 \right)
    T_F + \calO(\eps)
\right\},
\end{align}
with $L\equiv\ln\mu^2/m_b^2$ and the number of massless quarks $n_l=4$.

\subsection{Renormalized form factors}
\label{sec:renFF}

We now present our results for the renormalized form factors
$F_i(q^2)$. It should be noticed that these form factors are formally IR
divergent. We address the subtraction of these IR-divergences in the
following section. 

It will be convenient to further decompose the coefficients $F_i^{(k)}$
of the perturbative expansion (\ref{eq:pertexp}) according to 
\begin{align}
F_i^{(k)} = \sum_j \; F_{i,j}^{(k)} \; \eps^j.
\end{align}
We suppress an overall prefactor $i g_w V_{ub}/2\sqrt{2}$ and recall
that $q^2=\ub m_b^2$ with $\ub=1-u$ and $L=\ln\mu^2/m_b^2$. In our
normalization the form factors become in leading order 
\begin{align}
F_1^{(0)}(u) = 1,
\hspace{1.5cm}
F_2^{(0)}(u)  = 0,
\hspace{1.5cm}
F_3^{(0)}(u)  = 0.
\end{align}
In NLO we give our result up to terms of $\calO(\eps^2)$, which are
required for the IR-subtractions that we consider in the following
section. In terms of a set of coefficient functions $f_i(u)$,
$i=1,..,10$, which we list in Appendix~\ref{app:1loop}, we find that the
form factor $F_1$ is given by
\begin{align}
F_{1,-2}^{(1)}(u)  &= -C_F,
\no\\
F_{1,-1}^{(1)}(u)  &= C_F \bigg( f_1(u) - L\bigg),
\no\\
F_{1,0}^{(1)}(u)  &= C_F \bigg( f_2(u) + f_1(u) L - \frac12 L^2\bigg),
\no\\
F_{1,1}^{(1)}(u)  &= C_F \bigg( f_3(u) + f_2(u) L + \frac12
f_1(u) L^2 - \frac16 L^3\bigg),
\no\\
F_{1,2}^{(1)}(u)  &= C_F \bigg( f_4(u) + f_3(u) L + \frac12 f_2(u) L^2 +
\frac16 f_1(u) L^3 - \frac{1}{24} L^4\bigg),
\label{eq:F1NLO}
\end{align}
whereas the other two form factors are finite at NLO and read
\begin{align}
F_{2,0}^{(1)}(u)  &= C_F  f_5(u),
\no\\
F_{2,1}^{(1)}(u)  &= C_F \bigg( f_6(u) + f_5(u) L\bigg),
\no\\
F_{2,2}^{(1)}(u)  &= C_F \bigg( f_7(u) + f_6(u) L + \frac12 f_5(u)
L^2\bigg),
\label{eq:F2NLO}
\end{align}
and
\begin{align}
F_{3,0}^{(1)}(u)  &= C_F f_8(u),
\no\\
F_{3,1}^{(1)}(u)  &= C_F \bigg( f_9(u) + f_8(u) L\bigg),
\no\\
F_{3,2}^{(1)}(u)  &= C_F \bigg( f_{10}(u) + f_9(u) L + \frac12 f_8(u)
L^2\bigg). 
\label{eq:F23NLO}
\end{align}
In NNLO the divergent terms of the form factors can also be expressed in
terms of the 1-loop coefficient functions $f_i(u)$. The divergent terms
of the form factor $F_1$ read 
\begin{align}
F_{1,-4}^{(2)}(u)  &= \frac12 C_F^2,
\no\\
F_{1,-3}^{(2)}(u)  &= C_F^2 \bigg(L - f_1(u)\bigg) +
  \frac{11}{4} C_A C_F - n_l T_F C_F,
\no\\
F_{1,-2}^{(2)}(u)  &= C_F^2 \bigg[ L^2 - 2 f_1(u) L + \frac12 f_1(u)^2 -
f_2(u) \bigg]
+ C_A C_F \bigg[ \frac{11}{6} \bigg(L - f_1(u)\bigg)  - \frac{67}{36} +
\frac{\pi^2}{12}  \bigg] 
\no\\
& \quad
+ n_l T_F C_F \bigg[ \frac59 - \frac23 \bigg(L - f_1(u) \bigg)  \bigg] +
\frac43 L \, T_F C_F,
\no\\
F_{1,-1}^{(2)}(u)  &= C_F^2  \bigg[ \frac23 L^3 - 2 f_1(u) L^2 -
\bigg(2f_2(u)-f_1(u)^2\bigg) L + f_1(u) f_2(u) - f_3(u) - \frac38 +
\frac{\pi^2}{2}
\no\\
& \quad
- 6 \zeta_3 \bigg]
+ C_A C_F \bigg[ \bigg(\frac{\pi^2}{6} - \frac{67}{18} \bigg) \bigg(L -
f_1(u)\bigg) + \frac{461}{216} - \frac{17\pi^2}{24} + \frac{11}{2}
\zeta_3 \bigg]
\no\\
& \quad
+ n_l T_F C_F \bigg[ \frac{10}{9} \bigg(L - f_1(u)\bigg) -\frac{25}{54} +
\frac{\pi^2}{6} \bigg] + T_F C_F \bigg[ 2L^2 - \frac43 f_1(u) L +
\frac{\pi^2}{9} \bigg],
\label{eq:F1NNLO:div}
\end{align}
whereas the other two form factors start with a $1/\eps^2$-singularity,
\begin{align}
F_{2,-2}^{(2)}(u)  &= -C_F^2 f_5(u),
\no\\
F_{2,-1}^{(2)}(u)  &= C_F^2 \bigg( f_1(u) f_5(u) - f_6(u) - 2 f_5(u) L
\bigg),
\label{eq:F2NNLO:div}
\end{align}
and
\begin{align}
F_{3,-2}^{(2)}(u)  &= -C_F^2 f_8(u),
\no\\
F_{3,-1}^{(2)}(u)  &= C_F^2 \bigg( f_1(u) f_8(u) - f_9(u) - 2 f_8(u) L
\bigg). 
\label{eq:F23NNLO:div}
\end{align}
The finite terms of the 2-loop form factors give rise to a new set of
coefficient functions $k_i(u)$, $i=1,..,10$, which we collect in
Appendix~\ref{app:2loop}. We find 
\begin{align}
F_{1,0}^{(2)}(u) &= C_F^2 \bigg[ \frac13 L^4 - \frac43 f_1(u) L^3 -
\bigg(2f_2(u) - f_1(u)^2 \bigg) L^2 - \bigg(2f_3(u) - 2 f_2(u) f_1(u) +
\frac34
\no\\
& \quad
- \pi^2 + 12 \zeta_3 \bigg) L + k_1(u) \bigg]
+ C_A C_F \bigg[ - \frac{11}{18} L^3 + \bigg( \frac{11}{6} f_1(u)
-\frac{67}{18} + \frac{\pi^2}{6} \bigg)L^2
\no\\
& \quad
+ \bigg( \frac{11}{3} f_2(u) + \Big( \frac{67}{9} - \frac{\pi^2}{3} \Big)
f_1(u) + \frac{461}{108} - \frac{17\pi^2}{12} + 11 \zeta_3 \bigg)L +
k_2(u) \bigg]
\no\\
& \quad
+ n_l T_F C_F \bigg[ \frac29 L^3 - \bigg( \frac23 f_1(u) - \frac{10}{9}
\bigg)L^2 - \bigg(\frac43 f_2(u) + \frac{20}{9} f_1(u) + \frac{25}{27} -
\frac{\pi^2}{3} \bigg) L
\no\\
& \quad
- \frac43 f_3(u) - \frac{20}{9} f_2(u) + \bigg( \frac{25}{27} -
\frac{\pi^2}{3} \bigg) f_1(u) + k_3(u) \bigg]
\no\\
& \quad
+ T_F C_F \bigg[\frac{14}{9} L^3 - 2 f_1(u) L^2 - \bigg( \frac43
f_2(u) - \frac{2\pi^2}{9} \bigg)L + k_4(u) \bigg],
\no\\
F_{2,0}^{(2)}(u) &= C_F^2 \bigg[ k_5(u) + 2 \bigg( f_1(u) f_5(u) - f_6(u)
\bigg) L - 2 f_5(u) L^2 \bigg]
\!+\! C_A C_F \bigg[ k_6(u) + \frac{11}{3} f_5(u) L \bigg]
\no\\
& \quad
+ n_l T_F C_F \bigg[ -\frac43 f_6(u) - \frac{14}{9} f_5(u) - \frac43
f_5(u) L \bigg]
+ T_F C_F \bigg[ k_7(u) - \frac43 f_5(u) L \bigg],
\no\\
F_{3,0}^{(2)}(u) &= C_F^2 \bigg[ k_8(u) + 2 \bigg( f_1(u) f_8(u) - f_9(u)
\bigg) L - 2 f_8(u) L^2 \bigg]
\!+\! C_A C_F \bigg[ k_9(u) + \frac{11}{3} f_8(u) L \bigg]
\no\\
& \quad
+ n_l T_F C_F \bigg[ \frac{8}{3\ub}\ln(u) -\frac43 f_9(u) - \frac{14}{9}
f_8(u) - \frac43 f_8(u) L \bigg]
\no\\
& \quad
+ T_F C_F \bigg[ k_{10}(u) - \frac43 f_8(u) L \bigg].
\label{eq:F123NNLO:fin}
\end{align}
We compared our results for the 2-loop form factors with the ones given
in~\cite{Bonciani:2008wf} and found complete analytical agreement.


\section{Matching calculation}

\label{sec:matching}

\subsection{IR-subtractions}
\label{sec:IR}

In order to extract the Wilson coefficients $C_i$ from the form factors
$F_i$ of the previous section, we have to consider the perturbative
expansion of the hadronic tensor $W^{\mu\nu}$ and to absorb the
IR-singularities of the form factors into the jet function $J$ and the
shape function $S$. As the convolution of jet and shape function is
independent from the indices $(i,j)$ it is convenient to rewrite the
factorization formula (\ref{eq:ff:exact}) as $W^{\mu\nu} = \sum
c_{ij}^{\mu\nu} W_{ij}$ with 
\begin{align}
W_{i j} = \, C_i C_j \; J \otimes S,
\end{align}
where the symbol $\otimes$ represents the convolution integral. When we
write the perturbative expansion in terms of the renormalized coupling
constant, we should keep in mind that the actual 2-loop calculation has
been performed in QCD with five active quark flavours, while the NNLO
calculation of the jet function~\cite{Becher:2006qw} and the partonic
shape function~\cite{Becher:2005pd} have been performed in SCET with
four active flavours. We therefore have to account for an additional
finite renormalization to consistently express the perturbative
expansion in terms of the renormalized coupling constant of the
four-flavour theory, 
\begin{align}
\as^{(5)}(\mu) = \as^{(4)}(\mu) \left[ 1 + \frac{\as^{(4)}(\mu)}{4\pi}
  \,\delta\as^{(1)} + \calO(\as^2)  \right],
\label{eq:alphas5}
\end{align}
with (cf.~e.g.~\cite{Chetyrkin:1997un})
\begin{align}
\delta\as^{(1)} = T_F \bigg[ \frac43 L + \bigg( \frac23 L^2 +
\frac{\pi^2}{9} \bigg) \eps + \bigg( \frac29 L^3 + \frac{\pi^2}{9} L
-\frac49 \zeta_3 \bigg) \eps^2 + \calO(\eps^3) \bigg].
\end{align}
In the four-flavour theory the perturbative expansion of the hadronic
tensor becomes 
\begin{align}
W_{i j}^{(0)} &= \, C_i^{(0)} C_j^{(0)} \; \big[ J \otimes S \big]^{(0)},
\no\\
W_{i j}^{(1)} &= \, \left( C_i^{(1)} C_j^{(0)}  + C_i^{(0)} C_j^{(1)}
\right) \big[ J \otimes S \big]^{(0)} + C_i^{(0)} C_j^{(0)} \; \big[ J
\otimes S \big]^{(1)},
\no\\
W_{i j}^{(2)} &=
\, \left( C_i^{(2)} C_j^{(0)}  + C_i^{(1)} C_j^{(1)} + C_i^{(0)}
  C_j^{(2)} - \delta\as^{(1)} C_i^{(1)} C_j^{(0)}
  - \delta\as^{(1)} C_i^{(0)} C_j^{(1)} \right) \big[ J \otimes S
  \big]^{(0)}
\no\\
&\quad
+ \left( C_i^{(1)} C_j^{(0)}  + C_i^{(0)} C_j^{(1)} - \delta\as^{(1)}
  C_i^{(0)} C_j^{(0)}  \right) \big[ J \otimes S \big]^{(1)} + C_i^{(0)}
C_j^{(0)} \; \big[ J \otimes S \big]^{(2)}.
\end{align}
As we have mentioned in Section~\ref{sec:setup}, the matching
calculation simplifies in our setup due to the fact that the SCET
diagrams are scaleless and vanish. The IR-subtractions are therefore
entirely determined by the counterterms. At NLO the counterterm can be
inferred from~\cite{NLO}. In terms of the coefficient function $f_1(u)$
from Appendix~\ref{app:1loop}, it reads 
\begin{align}
\big[ J \otimes S \big]^{(1)} &=
C_F \left\{ -\frac{2}{\eps^2} - \frac{2}{\eps} \bigg( L - f_1(u) \bigg)
\right\} \;\big[ J \otimes S \big]^{(0)}.
\end{align}
In NNLO the counterterm can be extracted from the analysis
in~\cite{Becher:2006qw, Becher:2005pd}. We obtain 
\begin{align}
\big[ J \otimes S \big]^{(2)} &=
C_F \left\{ \frac{2C_F}{\eps^4}
+ \bigg[ 4 \bigg( L - f_1(u) \bigg) C_F + \frac{11}{2} C_A - 2 n_l T_F
\bigg]\frac{1}{\eps^3} \right.
\no\\
&\qquad
+ \bigg[ \bigg(2 L^2 - 4 f_1(u) L + 2f_1(u)^2\bigg) C_F + \bigg(
\frac{\pi^2}{6} - \frac{67}{18}  + \frac{11}{3} \Big( L - f_1(u) \Big)
\bigg) C_A
\no\\
&\qquad
+ \bigg( \frac{10}{9} - \frac{4}{3} \Big(L - f_1(u)\Big) \bigg) n_l T_F
\bigg] \frac{1}{\eps^2}
+ \bigg[ \bigg( \pi^2 - \frac{3}{4} - 12 \zeta_3 \bigg) C_F
\no\\
&\qquad
+ \bigg( \frac{461}{108} - \frac{17\pi^2}{12} + 11 \zeta_3 + \Big(
\frac{\pi^2}{3}  - \frac{67}{9} \Big) \Big( L - f_1(u) \Big) \bigg) C_A
\no\\
&\qquad
+ \bigg( \frac{\pi^2}{3}- \frac{25}{27} + \frac{20}{9} \Big(L - f_1(u)
\Big) \bigg) n_l T_F\bigg] \frac{1}{\eps} \bigg\}\;
\big[ J \otimes S \big]^{(0)}.
\end{align}

\subsection{Hard coefficient functions in NNLO}

We now have assembled all pieces required for the NNLO calculation of
the hard coefficient functions $H_{ij}$. Performing the IR-subtractions
as described in the previous section, we observe that the UV- and
IR-divergences cancel in the hard coefficient functions which represents
both a non-trivial confirmation of the factorization formula
(\ref{eq:ff:exact}) and a stringent cross-check of our calculation. 

We give the results for the hard coefficient functions in terms of the
Wilson coefficients $C_i$, which arise in the matching relation
(\ref{eq:matching}) of the heavy-to-light current, $H_{ij}=C_iC_j$. We
remind that the perturbative expansion is formulated in terms of the
renormalized coupling constant in a four-flavour theory and that in
leading order 
\begin{align}
C_1^{(0)}(u)=1,
\hspace{1.5cm}
C_2^{(0)}(u)=0,
\hspace{1.5cm}
C_3^{(0)}(u)=0.
\end{align}
In NLO we obtain in terms of our results from Section~\ref{sec:renFF},
\begin{align}
C_1^{(1)}(u) &= F_{1,0}^{(1)}(u),
\no\\
C_2^{(1)}(u) &= F_{2,0}^{(1)}(u),
\no\\
\frac{2}{u} C_3^{(1)}(u) &= F_{3,0}^{(1)}(u).
\label{eq:Ci:NLO}
\end{align}
The NNLO coefficient functions were so far unknown. They are found in
this work to be
\begin{align}
C_{1}^{(2)}(u) &= F_{1,0}^{(2)}
+ T_F C_F \bigg[ \frac{4}{9} \zeta_3 + \frac{\pi^2}{9} f_1(u)
+ \frac{2}{9} \Big( 6 f_2(u) - \pi^2 \Big) L
+ 2 f_1(u) L^2 - \frac{14}{9} L^3 \bigg]
\no\\
& \quad
+ C_F^2 \bigg[ f_4(u) - f_1(u) f_3(u) + \Big( 2 f_3(u) - f_1(u) f_2(u)
\Big) L
\no\\
& \hspace{1.5cm}
+ \frac12 \Big( 3 f_2(u) - f_1(u)^2 \Big) L^2
+ \frac56 f_1(u) L^3 - \frac{5}{24} L^4 \bigg],
\label{eq:C1:NNLO}
\end{align}
and
\begin{align}
C_{2}^{(2)}(u) &= F_{2,0}^{(2)}
+ T_F C_F \bigg[ \frac43 f_5(u) L \bigg]
\no\\
& \quad
+ C_F^2 \bigg[ f_7(u) - f_1(u) f_6(u)
+ \Big( 2f_6(u) - f_1(u) f_5(u) \Big) L
+ \frac32 f_5(u) L^2 \bigg],
\label{eq:C2:NNLO}
\end{align}
and
\begin{align}
\frac{2}{u} C_{3}^{(2)}(u) &= F_{3,0}^{(2)}
+ T_F C_F \bigg[ \frac43 f_8(u) L \bigg]
\no\\
& \quad
+ C_F^2 \bigg[ f_{10}(u) - f_1(u) f_9(u)
+ \Big( 2f_9(u) - f_1(u) f_8(u) \Big) L
+ \frac32 f_8(u) L^2 \bigg].
\label{eq:C3:NNLO}
\end{align}
The hard functions $H_{ij}$ and hence the Wilson coefficients $C_i$ are
renormalization scale dependent. According to the factorization formula
(\ref{eq:ff:exact}), this scale dependence has to cancel against the one
of the convolution of jet and shape function. The scale dependence of
the Wilson coefficients $C_i$ is governed by the renormalization group
equation, 
\begin{align}
\frac{d}{d\ln\mu} C_i(u;\mu) =
\bigg[ \Gamma_\text{cusp}(\as) \ln \frac{um_b}{\mu}
+ \gamma'(\as) \bigg] C_i(u;\mu),
\label{eq:RGE}
\end{align}
where $\Gamma_\text{cusp}$ is the universal cusp anomalous dimension
which describes the renormalization properties of light-like Wilson
lines~\cite{Korchemsky:1987wg}. Writing the perturbative expansion as  
\begin{align}
\Gamma_\text{cusp}(\as)=
\sum_{k=1}^\infty \left( \frac{\as}{4\pi} \right)^k
\Gamma_\text{cusp}^{(k)},
\hspace{1.5cm}
\gamma'(\as) =
\sum_{k=1}^\infty  \left( \frac{\as}{4\pi} \right)^k
\gamma'^{(k)},
\end{align}
we derive from our NLO expressions in (\ref{eq:Ci:NLO}) the familiar
results 
\begin{align}
\Gamma_\text{cusp}^{(1)} = 4 C_F,
\hspace{1.5cm}
\gamma'^{(1)} = -5 C_F.
\end{align}
Similarly, we can read off from our NNLO results (\ref{eq:C1:NNLO}) -
(\ref{eq:C3:NNLO}) that 
\begin{align}
\Gamma_\text{cusp}^{(2)} &=
C_A C_F \bigg[ \frac{268}{9} - \frac{4\pi^2}{3} \bigg]
- \frac{80}{9} n_l T_F C_F,
\no\\
\gamma'^{(2)} &=
C_F^2 \bigg[ 2 \pi^2 - \frac32  - 24 \zeta_3 \bigg]
+ C_A C_F \bigg[ 22 \zeta_3 - \frac{1549}{54} - \frac{7\pi^2}{6} \bigg]
+ n_l T_F C_F \bigg[ \frac{250}{27} + \frac{2\pi^2}{3} \bigg].
\end{align}
While our result for the 2-loop cusp anomalous dimension is in agreement
with~\cite{Korchemsky:1987wg}, the expression for $\gamma'^{(2)}$ has
not yet been computed directly so far. Our result confirms the
conjecture in~\cite{Neubert:2004dd}, which was based on known results of
the 2-loop anomalous dimensions of the jet and the shape function.


\section{Charm mass effects}

\label{sec:charm}

In this section we extend our analysis to include charm mass
effects. The charm quark enters the 2-loop calculation through the
1-loop gluon self energy in the last diagram of
Figure~\ref{fig:Diagrams}. So far we treated the charm quark as massless
and the respective effects are given in the above formulas by the terms
that are proportional to $n_l=4$, which denotes the number of massless
quarks. 

A comment is in order about the power counting of the charm quark mass
within the effective theory approach. One possible choice is given by
$m_c\sim\mu_{hc}\sim(\LQCD m_b)^{1/2}$, i.e.~the additional mass scale
is considered to be of the order of the intermediate hard-collinear
scale. In this case charm mass effects enter in the calculation of the
jet function. In this setup the charm mass represents an IR-scale in the
QCD$\to$SCET matching calculation and can therefore be set to zero. If
we adopt the scaling $m_c\sim\mu_{hc}$, the hard coefficient functions
$H_{ij}$ are thus given by the formulas from
Section~\ref{sec:matching}. 

A second choice, which has been widely applied in studies of $b\to c$
decays, is to consider the charm quark in the heavy quark limit with
$m_b\to\infty$ and $m_c\to\infty$ while $m_c/m_b$ is kept fixed. In this
scenario the hard coefficient functions $H_{ij}$ depend on the mass
ratio $z=m_c/m_b$. On the other hand the jet function does not dependent
on the charm quark mass in this setup, since it is defined in SCET with
three active quark flavours. Both approaches differ in the fact that the
first one allows to resum logarithms of the form $\ln m_c/m_b$ while the
second one does not.

In the following we consider the second scenario, i.e.~we keep a finite
charm quark mass in the 2-loop QCD calculation and the subsequent
matching calculation. We briefly outline the steps of the calculation
that change when we introduce a finite charm quark mass and write our
results schematically as (for an arbitrary quantity $\calQ$) 
\begin{align}
\calQ(z) = \hat{\calQ} + \Delta \calQ(z),
\end{align}
where $\hat{\calQ}$ refers to the respective expression from
Section~\ref{sec:QCD} or~\ref{sec:matching} but with $n_l=3$ massless
quark flavours and $\Delta \calQ(z)$ gives the additional contribution
from a massive charm quark.

\subsection{QCD form factors}

\begin{figure}[b!]
\centerline{\parbox{14cm}{\centerline{
\includegraphics[width=14cm]{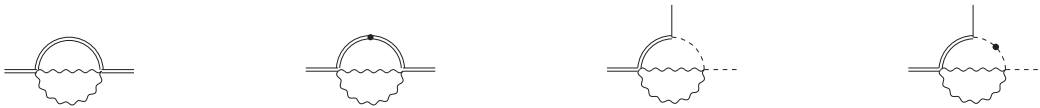}}
\caption{\label{fig:MIscharm} \small \textit{Additional Master Integrals
    with propagators of mass $m_c$ (wavy line). Notation as in
    Figure~\ref{fig:MIs}.}}}} 
\end{figure}

The charm mass enters the QCD calculation through the 2-loop diagram
with a closed charm quark loop, which gives rise to four additional MIs
(depicted in Figure~\ref{fig:MIscharm}). Since these MIs have not been
considered in~\cite{GB}, we present our results for these MIs in
Appendix~\ref{app:MIs}. As discussed in the appendix, we obtain
analytical results apart from the finite terms of the 4-topology MIs. We
could, however, find suitable representations which allow us to evaluate
these terms numerically to very high precision. Moreover, we provide
compact parameterizations for physical values of $z=m_c/m_b$ in the
appendix, which reproduce our numerical results at the percent level. 

The charm mass also modifies the renormalization procedure. We get an
additional contribution to the 2-loop wave-function renormalization
factor of the $u$-quark, 
\begin{align}
\Delta Z_{2,u}^{(2)}(z) = T_F C_F
\left( \frac{e^{\gamma_E}\mu^2}{z^2 m_b^2} \right)^{2\eps} \Gamma(\eps)^2 \;
\frac{2\eps(3-2\eps)(1+\eps)}{(1-\eps)(2-\eps)(1+2\eps)(3+2\eps)},
\end{align}
whereas the one of the $b$-quark receives the additional
contribution~\cite{Z2bOS:2loop}, 
\begin{align}
&\Delta Z_{2,b}^{(2)}(z) = T_F C_F
\bigg\{
\bigg( 1 + 4 L_c \bigg) \frac{1}{\eps} +6 L_c^2
+ \bigg( \frac{22}{3} + 8 \ln (z) \bigg)\! L_c
+ \frac{(5-18z-30z^3+12z^4)\pi^2}{3}
\no\\
&\qquad
+ \frac{4}{3} (19+12z^2) \ln (z)
+ 8 (1+3z^4) \ln^2 (z)
- 8z(3+5z^2) \bigg( \Li_2(-z) + \ln (z)  \ln(1+z) \bigg)
\no\\
&\qquad
- 2(1-z)(2-z-z^2-6z^3) \bigg( \Li_2(z^2) + 2\ln (z) \ln(1-z^2) \bigg)
+ \frac{443}{18} + 28 z^2
+ \calO(\eps)
\bigg\},
\end{align}
where we introduced $L_c = \ln \mu^2/m_c^2 = L - 2 \ln(z)$.

These changes modify our NNLO expressions for the renormalized QCD form
factors $F_i(q^2)$ from Section~\ref{sec:renFF}. We now introduce three
additional coefficient functions $k_{i}(u,z), i=11,..,13$, which we list in
Appendix~\ref{app:2loop}. The form factor $F_1$ receives the additional
contribution  
\begin{align}
\Delta F_{1,-2}^{(2)}(u,z) &= \frac43 L_c T_F C_F,
\no\\
\Delta F_{1,-1}^{(2)}(u,z) &= \bigg[ 2L_c^2 - \frac43 \Big( f_1(u) -2 \ln (z)
\Big) L_c + \frac{\pi^2}{9} \bigg] T_F C_F,
\no\\
\Delta F_{1,-0}^{(2)}(u,z) &= \bigg[ \frac{14}{9} L_c^3
- 2 \Big( f_1(u) -2 \ln (z) \Big)L_c^2
\no\\
&\qquad
- \frac43 \bigg( f_2(u) + 2 \ln(z) f_1(u) - 2\ln^2 (z) - \frac{\pi^2}{6}
\bigg) L_c
+ k_{11}(u,z) \bigg] T_F C_F,
\label{eq:F1NNLO:charm}
\end{align}
whereas the other two form factors are modified by
\begin{align}
\Delta F_{2,0}^{(2)}(u,z) &= \bigg[ k_{12}(u,z) - \frac43 f_5(u) L_c \bigg]
T_F C_F,
\no\\
\Delta F_{3,0}^{(2)}(u,z) &= \bigg[ k_{13}(u,z) - \frac43 f_8(u) L_c \bigg]
T_F C_F.
\label{eq:F23NNLO:charm}
\end{align}

\subsection{Hard coefficient functions}

According to the power counting that we adopt for the charm quark mass
scale, the charm quark is integrated out in the QCD$\to$SCET matching
calculation. The IR-subtractions turn out to be analogous to those of
Section~\ref{sec:IR}, apart from the fact that we now match onto SCET
with three active quark flavours. If we express the perturbative
expansion in terms of the renormalized coupling constant of the
three-flavour theory, we thus have to account for an additional
contribution to (\ref{eq:alphas5}) with
\begin{align}
\Delta \delta \as^{(1)}(z) =
T_F \bigg[ \frac43 L_c + \bigg( \frac23 L_c^2 +
\frac{\pi^2}{9} \bigg) \eps + \bigg( \frac29 L_c^3 + \frac{\pi^2}{9} L_c
-\frac49 \zeta_3 \bigg) \eps^2 + \calO(\eps^3) \bigg],
\end{align}
where we recall that $L_c = \ln \mu^2/m_c^2 = L - 2 \ln(z)$.

We summarize our results for the hard coefficient functions $H_{ij}=C_i
C_j$. If we keep a finite charm quark mass in the matching calculation,
which corresponds to the power counting $m_b\to\infty$ and
$m_c\to\infty$ with $m_c/m_b$ fixed, we find that the coefficients
functions are given in NNLO by our results from (\ref{eq:C1:NNLO}) -
(\ref{eq:C3:NNLO}) with $n_l=3$ massless quarks and the additional
contributions 
\begin{align}
\Delta C_1^{(2)}(u,z) &= \bigg[ k_{11}(u,z) + \bigg( f_1(u) - 2 \ln(z)
\bigg) \frac{\pi^2}{9} + \frac49 \zeta_3 \bigg]T_F C_F,
\no\\
\Delta C_2^{(2)}(u,z) &= k_{12}(u,z) \;T_F C_F,
\no\\
\frac{2}{u}\Delta C_3^{(2)}(u,z) &= k_{13}(u,z) \;T_F C_F,
\end{align}
which depend on the quark mass ratio $z=m_c/m_b$.


\section{Numerical analysis}

\label{sec:numerics}

\begin{figure}[ht!]
\centerline{\parbox{14cm}{\centerline{
 \psfrag{u}{$u$}
 \psfrag{C1}{$C_1(u)$}
 \includegraphics[width=9.2cm]{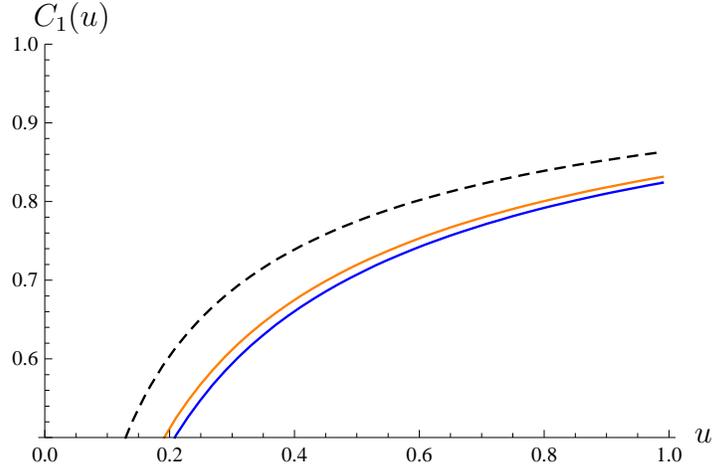}}
\vspace{11mm}
\centerline{
 \psfrag{u}{$u$}
 \psfrag{C2}{$C_2(u)$}
 \includegraphics[width=9.2cm]{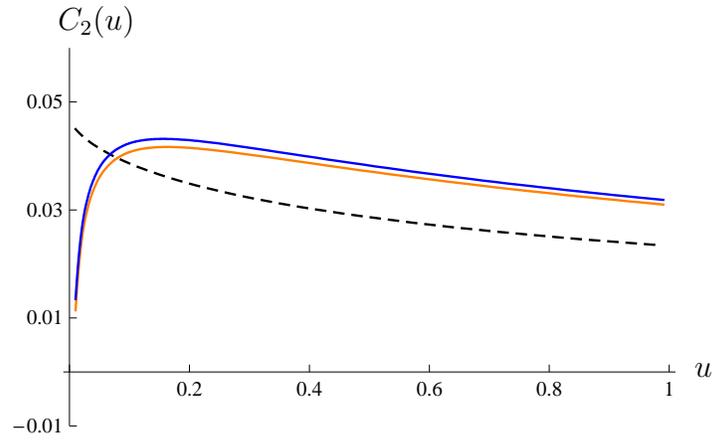}}
\vspace{11mm}
\centerline{
 \psfrag{u}{$u$}
 \psfrag{C3}{$C_3(u)$}
 \includegraphics[width=9.2cm]{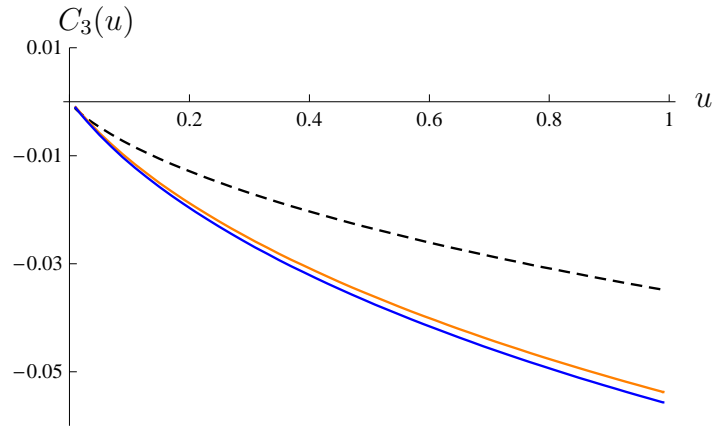}}
\caption{\label{fig:Cis} \small \textit{Wilson coefficients $C_i$ at the
    scale $\mu=m_b$ as a function of $u$ (the momentum transfer is given
    by $q^2=(1-u)m_b^2$). The dashed lines refer to the NLO results and
    the solid lines to the NNLO results with $z=m_c/m_b=0$ (orange/light
    gray) and $z=0.3$ (blue/dark gray).}}}}
\end{figure}

\begin{figure}[ht!]
\centerline{\parbox{14cm}{\centerline{
 \psfrag{u}{$u$}
 \psfrag{C1}{$C_1^{(2)}(u)$}
 \includegraphics[width=9.2cm]{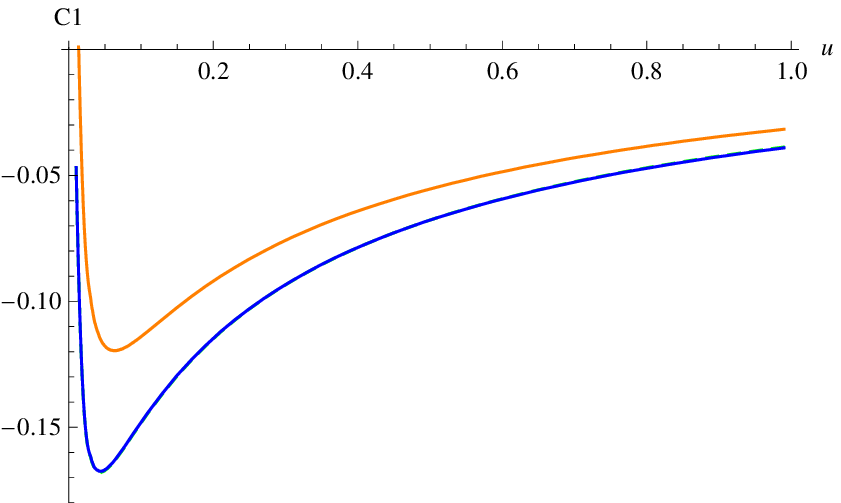}}
\vspace{11mm}
\centerline{
 \psfrag{u}{$u$}
 \psfrag{C2}{$C_2^{(2)}(u)$}
 \includegraphics[width=9.2cm]{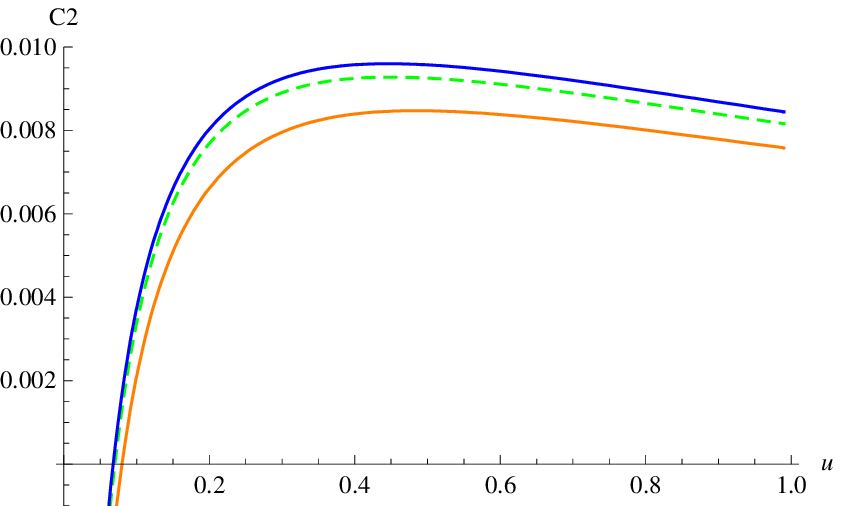}}
\vspace{11mm}
\centerline{
 \psfrag{u}{$u$}
 \psfrag{C3}{$C_3^{(2)}(u)$}
 \includegraphics[width=9cm]{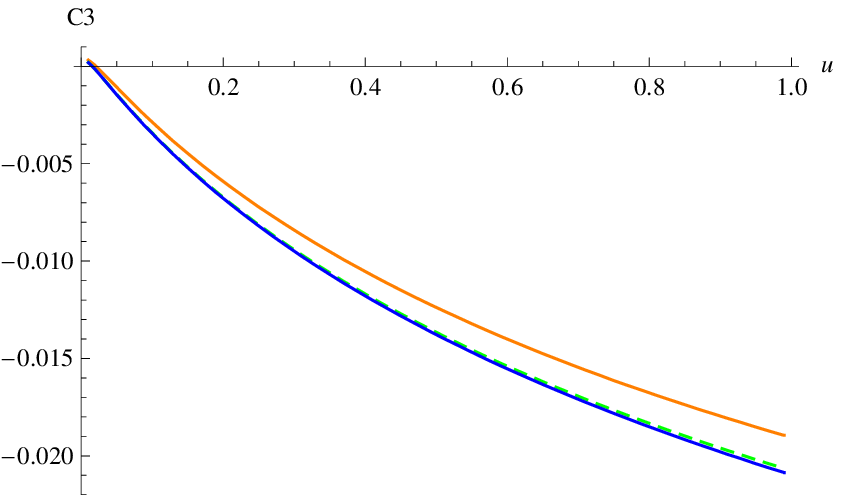}}
\caption{\label{fig:Cis:NNLO}\small \textit{NNLO contributions to the
    Wilson coefficients $C_i$ at the scale $\mu=m_b$ as a function of
    $u$. The solid lines again refer to the NNLO results with 
    $z=m_c/m_b=0$ (orange/light gray) and $z=0.3$ (blue/dark gray). The
    dashed green/light gray line shows the limit $z=1$. In the first and
    the third plot the dashed line can hardly be distinguished from the
    solid blue/dark gray line.}}}} 
\end{figure}

We briefly discuss the numerical impact of the considered NNLO
corrections. As a phenomenological analysis of inclusive semileptonic
$B$ decays is beyond the scope of the present paper, we illustrate our
results at the level of the Wilson coefficients $C_i$ which arise in the
matching relation (\ref{eq:matching}) of the semileptonic current
(recall that the hard coefficient functions from (\ref{eq:ff:exact}) are
given by $H_{ij}=C_iC_j$). We in particular do not study the
renormalization scale dependence, since the Wilson coefficients are
explicitly scale dependent, cf.~(\ref{eq:RGE}). The scale dependence is
cancelled only at the level of the hadronic tensor against the one of
the jet and the shape function. 

In Figure~\ref{fig:Cis} we show the Wilson coefficients $C_i$ at the
scale $\mu=m_b$. We see that the considered NNLO corrections give
moderate contributions to the Wilson coefficients, which add
constructively to the NLO corrections. The effect of the finite charm
quark mass is small, but amounts to $\sim10-20\%$ of the NNLO contribution
as can be seen in Figure~\ref{fig:Cis:NNLO}. In
Figure~\ref{fig:Cis:NNLO} we also illustrate a scenario, which considers
the charm quark to be as heavy as the $b$-quark, i.e.~$z=m_c/m_b=1$ (dashed
line). Somewhat surprising, the curves for physical charm masses with
$z\simeq0.3$ are much closer to the limit $m_c=m_b$ than to the limit
$m_c=0$ (notice that the effect is even slightly larger for physical 
charm masses than for $m_c=m_b$).


\section{Conclusion}

\label{sec:concl}

We computed NNLO corrections to the hard coefficient functions $H_{ij}$
which arise in the factorization formula (\ref{eq:ff:exact}) for
inclusive semileptonic $B$ meson decays in the shape-function
region. Together with the 2-loop corrections to the jet function
from~\cite{Becher:2006qw}, our calculation completes the NNLO
calculation of $B\to X_u\ell\nu$ decays in the shape-function region.  

The considered calculation is closely related to the 2-loop calculation
in charmless hadronic $B$ decays~\cite{GB}. In particular, all Master
Integrals that appear in the present calculation have already been
calculated in~\cite{GB}. We confirmed analytical results for the 2-loop
form factors which have been computed recently in~\cite{Bonciani:2008wf}
and extended the analysis to extract the hard coefficient functions
$H_{ij}$ from the form factors which are formally IR-divergent. We in
addition kept a finite charm quark mass in the 2-loop calculation.  

A phenomenological analysis of partial decay rates in the shape-function
region was beyond the scope of the present paper. We briefly discussed
the numerical impact of the considered corrections to the short-distance
coefficients, which was found to be moderate. One should keep in mind,
however, that the present calculation reflects only a part of the full
NNLO calculation.  We in particular expect that the renormalization
scale dependence of the theoretical prediction for partial decay
rates will be significantly reduced, once the renormalization group
improvement is taken into account in NNLO.

\vspace{2mm}

\textbf{Note added:} While this paper was in preparation, the work
in~\cite{Asatrian:2008uk,Beneke:2008ei} on the same topic
appeared. While the analysis in~\cite{Beneke:2008ei} did not go beyond
the one in~\cite{Bonciani:2008wf}, the authors of~\cite{Asatrian:2008uk}
also performed the full matching calculation\footnote{The revised
  version of \cite{Beneke:2008ei} now contains the full matching
  calculation.}. We have compared our results for the NNLO hard
coefficient functions from (\ref{eq:C1:NNLO}) - (\ref{eq:C3:NNLO}) with
those in~\cite{Asatrian:2008uk} and found agreement. None of the works
in~\cite{Asatrian:2008uk,Beneke:2008ei} addressed charm mass effects.


\subsection*{Acknowledgements}


We are grateful to Gerhard Buchalla for many interesting discussions and
comments on the manuscript. We thank Roberto Bonciani and Andrea
Ferroglia for discussions related to the 6-topology Master
Integral. This work was supported by the DFG
Sonder-forschungsbereich/Transregio 9.

\newpage
\begin{appendix}

\section{NLO coefficient functions}
\label{app:1loop}

The 1-loop calculation of the QCD form factors gives rise to the
following coefficient functions
\begin{align}
f_1(u) &= - \frac52 + 2 \ln (u),
\no\\
f_2(u) &=  - 6 - \frac{\pi^2}{12} - \frac{1-3\ub}{\ub} \ln (u) -
2 \ln^2(u) - 2 \Li_2(\ub),
\no\\
f_3(u) &= - 12 - \frac{5\pi^2}{24} + \frac13 \zeta_3 -
\frac{24(1-2\ub)-\pi^2\ub}{6\ub} \ln(u) + \frac{1-3\ub}{\ub}
\Big( \ln^2(u) + \Li_2(\ub) \Big)
\no\\
&\quad
+ \frac43 \ln^3(u) + 4 \ln(u) \Li_2(\ub) - 2 \Li_3(\ub) + 4
\S_{1,2}(\ub),
\no\\
f_4(u) &= - 24 - \frac{\pi^2}{2} - \frac{\pi^4}{160} + \frac56 \zeta_3 -
\frac{96(1-2\ub)+(1-3\ub)\pi^2+8\ub \zeta_3}{12\ub} \ln(u) 
\no\\
&\quad
+ \frac{24(1-2\ub)-\pi^2\ub}{6\ub} \Big( \ln^2(u) + \Li_2(\ub) \Big) -
\frac23 \ln^4(u) - 4 \ln^2(u) \Li_2(\ub) 
\no\\
&\quad
- \frac{1-3\ub}{\ub} \bigg( \frac23 \ln^3(u) + 2 \ln(u) \Li_2(\ub) -
\Li_3(\ub) + 2 \S_{1,2}(\ub) \bigg) - 8 \ln(u) \S_{1,2}(\ub) 
\no\\
&\quad
+ 4 \ln(u) \Li_3(\ub) - 2 \Li_4(\ub) - 8 \S_{1,3}(\ub) + 4 \S_{2,2}(\ub),
\no\\
f_5(u) &= \frac{2}{\ub} + \frac{2u}{\ub^2} \ln(u),
\no\\
f_6(u) &= \frac{4}{\ub} + \frac{2u}{\ub^2} \Big( \ln(u) - \ln^2(u) -
\Li_2(\ub) \Big), 
\no\\
f_7(u) &= \frac{48+\pi^2}{6\ub} + \frac{2u}{\ub^2} \bigg(
\frac{24+\pi^2}{12} \ln(u) - \ln^2(u) - \Li_2(\ub) + \frac23 \ln^3(u) +
2 \ln(u) \Li_2(\ub) 
\no\\
&\quad
- \Li_3(\ub) + 2 \S_{1,2}(\ub) \bigg),
\no\\
f_8(u) &= - \frac{2}{\ub} - \frac{2(1-2\ub)}{\ub^2} \ln(u),
\no\\
f_9(u) &= - \frac{4}{\ub} - \frac{2(1-5\ub)}{\ub^2} \ln(u) +
\frac{2(1-2\ub)}{\ub^2} \Big( \ln^2(u) + \Li_2(\ub) \Big), 
\no\\
f_{10}(u) &= - \frac{48+\pi^2}{6\ub} -
\frac{24(1-5\ub)+\pi^2(1-2\ub)}{6\ub^2} \ln(u) + \frac{2(1-5\ub)}{\ub^2}
\Big(\ln^2(u)+\Li_2(\ub)\Big) 
\no\\
&\quad
- \frac{2(1-2\ub)}{\ub^2} \bigg( \frac23 \ln^3(u) +2 \ln(u) \Li_2(\ub) -
\Li_3(\ub) + 2\S_{1,2}(\ub) \bigg). 
\end{align}

\section{NNLO coefficient functions}
\label{app:2loop}

In the 2-loop expressions of the QCD form factors
(\ref{eq:F123NNLO:fin}) we introduced the coefficient functions
$k_i(u)$, which read 
\begin{align}
k_1(u) &=
\frac{2(5 + 31\ub + 3\ub^2 + 3\ub^3)}{u^3} \Li_4(\ub)
+ \frac{4(13 + 53\ub + 15\ub^2 + 3\ub^3)}{u^3} \S_{2,2}(\ub)
+ 8 \S_{1,3}(\ub)
\no\\
&
- \frac{2(3 + 14\ub + 3\ub^2 + \ub^3)}{u^3} \Li_2(\ub)^2
- 8 \ln(u) \Li_3(\ub)
+ 16 \ln(u)\S_{1, 2}(\ub)
+ \frac{16}{3} \ln^4(u)
\no\\
&
+ 16 \ln^2(u)\Li_2(\ub)
+ \frac{3 + 100\ub + 145\ub^2 + 4\ub^3}{3u^2\ub} \Li_3(\ub)
+ \frac{42 - 68\ub}{9\ub} \ln^3(u)
\no\\
&
- \frac{2(3 - 242\ub - 233\ub^2 - 32\ub^3)}{3u^2\ub} \S_{1, 2}(\ub)
+ \frac{2(6 + 41\ub + 77\ub^2 + 2\ub^3)}{3u^2\ub} \ln(u) \Li_2(\ub)
\no\\
&
- \bigg(\frac{3 + 89\ub + 785\ub^2 + 635\ub^3}{18u^2\ub} + \frac{(5 +
  31\ub + 3\ub^2 + 3\ub^3)\pi^2}{u^3}\bigg) \Li_2(\ub) 
\no\\
&
- \bigg( \frac{147 - 736\ub - 167\ub^2}{18u\ub} - 2 \pi^2\bigg) \ln^2(u)
- \frac{(5267 + 10490 \ub + 2387 \ub^2)\pi^2}{432 u^2} 
\no\\
&
- \bigg( \frac{291 - 959\ub}{54\ub} - \frac{(9 + 340\ub + 355\ub^2 +
  52\ub^3) \pi^2}{18u^2\ub} + \frac{56}{3}\zeta_3\bigg) \ln(u) 
\no\\
&
+ \frac{(787 + 2607\ub + 1065\ub^2 + 77\ub^3)\pi^4}{720u^3} +
\frac{67}{9} \zeta_3 - \frac{71387}{1296} - 2 k_2(u), 
\no\\
k_2(u) &=
-\frac{4(1 - 3\ub + \ub^2 + 5\ub^3)}{12u^2\ub}
\Big( 12\calH_1(\ub) + \pi^2 \ln(2-u)\Big)
+ \frac{8(1 + 2\ub + 4\ub^2)}{u^3} \S_{2,2}(\ub)
\no\\
&
- \frac{2\ub - u^3}{3u^3} \Big( 24 \calH_2(\ub)- 2\pi^2 \Li_2(-\ub)\Big)
+ \frac{2(1 + 10\ub + 4\ub^2)}{u^3} \Li_4(\ub)
- 8 \ln(u)\Li_3(\ub)
\no\\
&
- \frac{5 - 10\ub + 16\ub^2 - 4\ub^3}{u^3} \Li_2(\ub)^2
+ \frac{8 + 36\ub - 33\ub^2 - 20\ub^3}{3u^3} \Li_3(\ub)
\no\\
&
+ \frac{92 - 24\ub - 33\ub^2 - 26\ub^3}{3u^3} \S_{1,2}(\ub)
+ \frac{47 - 49\ub + 44\ub^2}{3u^2} \ln(u)\Li_2(\ub)
+ \frac{44}{9} \ln^3(u)
\no\\
&
+ \bigg( \frac{33 - 227\ub + 157\ub^2 - 116\ub^3}{9u^2\ub} + \frac{(1 -
  26\ub - 4\ub^3) \pi^2}{3u^3}\bigg) \Li_2(\ub) 
\no\\
&
+ \frac{39 - 235\ub + 349\ub^2 + 12u \ub \pi^2}{18u \ub} \ln^2(u)
- \frac{(419 + 1178\ub + 815\ub^2)\pi^2}{216u^2}
\no\\
&
- \bigg( \frac{807 - 2545\ub}{54\ub} - \frac{(12 + 65\ub + 41\ub^2 +
  56\ub^3) \pi^2}{18u^2\ub} + 14 \zeta_3\bigg) \ln(u) 
+ \frac{289 - 19\ub}{18u} \zeta_3
\no\\
&
+ \frac{u^2 - 3}{u^3} \bigg( \Li_3(-u) - \ln(u)\Li_2(-u) -
\frac{\ln^2(u)+\pi^2}{2}\ln(1+u)\bigg) 
- \frac{89437}{1296}
\no\\
&
- \frac{(227 - 1935\ub + 519\ub^2 - 389\ub^3)\pi^4}{1080u^3} - \frac{(6
  + 2\ub)\pi^2}{u} \ln(2) - \calC_0, 
\no\\
k_3(u) &= -\frac{10}{3\ub} \ln(u) - \frac{2125}{324} + \frac{5\pi^2}{18}
+ \frac{10}{3} \zeta_3, 
\no\\
k_4(u) &=
\frac{8(6\ub + u^3)}{3u^3} \Li_3(\ub)
- \frac{4(3 + 8\ub - 24\ub^3 - 19\ub^4)}{9u^3\ub} \Li_2(\ub)
- \frac{144\ub + 28u^3}{9u^3} \zeta_3
\no\\
&
+ \frac{2(57 + 89 \ub - 73 \ub^2 - 265 \ub^3 - 3\ub u^2
  \pi^2)}{27u^2\ub} \ln(u) 
\no\\
&
+ \frac{(23 + 315\ub - 507 \ub^2 + 41 \ub^3)\pi^2}{54u^3}
+ \frac{1111 - 6758\ub + 7951\ub^2}{162u^2},
\no\\
k_5(u) &=
\frac{4(13 + 8\ub)}{u^3} \Big( 2 \Li_4(\ub) + 8 \S_{2,2}(\ub) -
\Li_2(\ub)^2 - \pi^2\Li_2(\ub) \Big) 
+ \frac{262 - 298\ub}{9\ub^2}\ln(u)
\no\\
&
- \frac{1 - 15\ub - 67\ub^2 -3\ub^3}{u^2\ub^2} \Big( 2\Li_3(\ub) + \pi^2
\ln(u)\Big) 
+ \frac{4(1 + 21\ub + 143\ub^2 + 3\ub^3)}{u^2\ub^2}\S_{1,2}(\ub)
\no\\
&
- \frac{4(2 - 12\ub - 29\ub^2 - 3\ub^3)}{u^2\ub^2}\ln(u)\Li_2(\ub)
- \frac{5 - 171\ub + 507\ub^2 + 163\ub^3}{3u^2\ub^2}\Li_2(\ub)
\no\\
&
- \frac{28u}{3\ub^2}\ln^3(u)
+ \frac{12 - 41\ub + 190\ub^2 + 91\ub^3}{3u \ub^3} \ln^2(u)
- \frac{(3 + 214\ub + 35\ub^2)\pi^2}{3u^2\ub}
\no\\
&
+ \frac{3(13 + 8\ub)\pi^4}{5u^3}
+ \frac{259}{9\ub}
- 2 k_6(u),
\no\\
k_6(u) &=
\frac{2(1 - 3\ub - 3\ub^2 + \ub^3)}{3u^2\ub^2} \Big( 12\calH_1(\ub) +
\pi^2 \ln(2-u)\Big) 
- \frac{4}{3u^3} \Big( 24\calH_2(\ub) - 2\pi^2 \Li_2(-\ub)\Big)
\no\\
&
+ \frac{4(11 + 4\ub)}{u^3} \Li_4(\ub)
+ \frac{2(3 + 4\ub)}{u^3} \Big(8\S_{2,2}(\ub) - \Li_2(\ub)^2\Big)
- \frac{13 - 56\ub - 8\ub^2}{3u \ub^2}\ln^2(u)
\no\\
&
+ \frac{2(2 + 30\ub - 38\ub^2 + 3\ub^3)}{u^3\ub} \Li_3(\ub)
+ \frac{2(8 + 36\ub - 30\ub^2 - 11\ub^3)}{u^3\ub} \S_{1,2}(\ub)
\no\\
&
+ \frac{2(2 + 9\ub + 3\ub^2)}{u^2\ub} \ln(u)\Li_2(\ub)
- \frac{(6 + 53\ub + 8\ub^2)\pi^2}{3u^2\ub}
+ \frac{269}{9\ub}
\no\\
&
- \Big(\frac{2(11 - 45\ub + 60\ub^2 + 25\ub^3)}{3u^2\ub^2}
+ \frac{2(17 + 12\ub) \pi^2}{3u^3}\Big)\Li_2(\ub)
- \frac{2(7 + \ub)}{u^2} \zeta_3
\no\\
&
+ \Big(\frac{203 - 257\ub}{9\ub^2} - \frac{(4 - 18\ub - 39\ub^2 -
  5\ub^3)\pi^2}{3u^2\ub^2}\Big)\ln(u) 
+ \frac{(155 +108\ub)\pi^4}{90u^3}
\no\\
&
- \frac{2(8 - 6\ub + \ub^2)}{u^3} \Big( \Li_3(-u) - \ln(u) \Li_2(-u) -
\frac{\ln^2(u)+\pi^2}{2}\ln(1+u)\Big), 
\no\\
k_7(u) &=
\frac{32}{u^3} \Li_3(\ub)
+ \frac{8(1 - 3\ub - 3\ub^2 + \ub^3)}{3u^2\ub^2} \Li_2(\ub)
- \frac{4(13 - 86\ub + 13\ub^2)}{9u \ub^2} \ln(u)
\no\\
&
+ \frac{32(3 - \ub) \pi^2}{9u^2}
- \frac{32}{u^3} \zeta_3
- \frac{4(19 + 59\ub)}{9u \ub},
\no\\
k_8(u) &=
- \frac{4(19 + 32\ub + 12\ub^2)}{u^4} \Big( 2 \Li_4(\ub) + 8
\S_{2,2}(\ub) - \Li_2(\ub)^2 - \pi^2\Li_2(\ub)\Big) 
\no\\
&
+ \frac{1 - 17\ub - 129\ub^2 - 101\ub^3 - 6\ub^4}{u^3\ub^2}\Big( 2
\Li_3(\ub) + \pi^2\ln(u)\Big) 
+ \frac{28(1 - 2\ub)}{3\ub^2}\ln^3(u)
\no\\
&
- \frac{4(1 + 19\ub + 285\ub^2 + 181\ub^3 + 18\ub^4)}{u^3\ub^2}
\S_{1,2}(\ub) 
- \frac{3(19 + 32\ub + 12\ub^2)\pi^4}{5u^4}
\no\\
&
+ \frac{4(2 - 16\ub - 51\ub^2 - 61\ub^3)}{u^3\ub^2}\ln(u)\Li_2(\ub)
+ \frac{(3 + 323\ub + 373\ub^2 + 57\ub^3)\pi^2}{3u^3\ub}
\no\\
&
+ \frac{5 - 187\ub + 363\ub^2 + 1123\ub^3 + 208\ub^4}{3u^3\ub^2}\Li_2(\ub)
- \frac{262 - 771\ub + 149\ub^2}{9u \ub^2}\ln(u)
\no\\
&
- \frac{12 - 53\ub + 182\ub^2 + 587\ub^3 + 28\ub^4}{3u^2\ub^3}\ln^2(u)
- \frac{259 - 151\ub}{9u \ub}
- 2k_9(u),\no\\
k_9(u) &=
- \frac{2(1 - 5\ub - 5\ub^2 - \ub^3 -
  2\ub^4)}{3u^3\ub^2}\Big(12\calH_1(\ub) + \pi^2 \ln(2 - u)\Big) 
\no\\
&
+ \frac{4(1 + 2\ub)}{3u^4}\Big( 24 \calH_2(\ub) - 2 \pi^2\Li_2(-\ub)\Big)
- \frac{4(13 + 24\ub + 8\ub^2)}{u^4} \Li_4(\ub)
\no\\
&
- \frac{2(5 + 8\ub + 8\ub^2)}{u^4} \Big( 8\S_{2,2}(\ub) - \Li_2(\ub)^2\Big)
- \frac{2(2 + 48\ub - 28\ub^2 - 25\ub^3 - 6\ub^4)}{u^4\ub} \Li_3(\ub)
\no\\
&
- \frac{2(8 + 70\ub + 16\ub^2 - 71\ub^3 - 14\ub^4)}{u^4\ub}\S_{1,2}(\ub)
+ \frac{13 - 82\ub - 34\ub^2 - 50\ub^3}{3u^2\ub^2}\ln^2(u)
\no\\
&
- \frac{2(2 + 17\ub + 17\ub^2 + 6\ub^3)}{u^3\ub}\ln(u)\Li_2(\ub)
+ \frac{(6 + 103\ub + 60\ub^2 + 32\ub^3)\pi^2}{3u^3\ub}
\no\\
&
+ \Big(\frac{2(11 - 67\ub + 72\ub^2 + 103\ub^3 + 34\ub^4)}{3u^3\ub^2}+
\frac{2(23 + 40\ub + 24\ub^2)\pi^2}{3u^4} \Big)\Li_2(\ub) 
\no\\
&
- \Big( \frac{203 - 795\ub + 430\ub^2}{9u \ub^2} - \frac{(4 - 26\ub -
  47\ub^2 - 103\ub^3 - 2\ub^4)\pi^2}{3u^3\ub^2}\Big)\ln(u) 
\no\\
&
+ \frac{2(22 - 20\ub + 9\ub^2 - 2\ub^3)}{u^4} \Big(\Li_3(-u) -
\ln(u)\Li_2(-u) - \frac{\ln^2(u)+\pi^2}{2}\ln(1+u) \Big) 
\no\\
&
+ \frac{2(23 - 17\ub + 10\ub^2)}{u^3} \zeta_3
- \frac{(209 + 364\ub +216\ub^2)\pi^4}{90u^4}
- \frac{8\pi^2}{u} \ln(2)
- \frac{269 - 215\ub}{9u \ub},
\no\\
k_{10}(u) &=
- \frac{32(1 + 2\ub)}{u^4} \Big(\Li_3(\ub) - \zeta_3\Big)
- \frac{8(1 - 5\ub - 21\ub^2 - 17\ub^3 - 2\ub^4)}{3u^3\ub^2} \Li_2(\ub)
\no\\
&
+ \frac{4(13 - 124\ub - 223\ub^2 - 38\ub^3)}{9u^2\ub^2} \ln(u)
- \frac{16(15 + 4\ub + \ub^2)\pi^2}{9u^3}
\no\\
&
+ \frac{4(19 + 364\ub + 43\ub^2)}{9u^2\ub}.
\end{align}
The definition of the functions $\calH_{1,2}(x)$ and the constant
$\calC_0$ can be found in Section~\ref{sec:2loop}. 

\newpage
When we include a finite charm quark mass in our calculation, we
encounter three additional coefficient functions which depend on the
quark mass ratio $z=m_c/m_b$. They have been introduced
in~(\ref{eq:F1NNLO:charm}) and~(\ref{eq:F23NNLO:charm}) and read 
\begin{align}
&k_{11}(u, z) =
\frac{4(u^3 + 6\ub z^4)}{3u^2z^2} \calI_1(\ub, z)
+ \frac{4(3u^4 - (3 - 19\ub)u^2 z^2 - 4(3 - 7\ub)\ub z^4)}{9u(1 +
  \ub)z^2} \calI_2(\ub,z) 
\no\\
&\quad
+ \bigg( \frac{24u^3(1-z) - (49 + 119\ub + 175\ub^2 - 55\ub^3) z^5 -
  18(1 + \ub)(1 - 8\ub + 3\ub^2) z^6}{9u^2(1 + \ub)z^2} 
\no\\
&\quad
+\frac{ (69 - 80\ub + 27\ub^2)u z^3 +4(15 - 2\ub + 23\ub^2)z^4  -2(31 -
  13\ub)u^2 z^2 }{9u^2(1 + \ub)z^2} \bigg) 
\no\\
&\quad
\times \bigg( \Li_2(z^2) + 2 \ln(z) \ln(1 - z^2) \bigg) 
+ \frac{4((3 - 19\ub)u^2 + 2(15 - 23\ub)\ub z^2)}{9u^2 \ub} \bigg(
\Li_2(\ub) + \ln^2(u) \bigg) 
\no\\
&\quad
+ \frac{4(24u^3 - (69 - 80\ub + 27\ub^2)u z^2 + (49 + 119\ub + 175\ub^2
  -55\ub^3) z^4)}{9u^2(1 + \ub)z} 
\no\\
&\quad
\times \bigg( \Li_2(-z) + \ln(z) \ln(1 + z) \bigg) 
- \frac{2(54u^3 - (21 - (37 + 3\pi^2)\ub)u z^2 + 216\ub z^4)}{27u \ub
  z^2} \ln(u) 
\no\\
&\quad
- \frac{4(12u^4 - (37 - 51\ub)u^2 z^2 + 2(15 - 32u\ub + \ub^2) z^4 -
  9(1 - 8\ub + 3\ub^2)(1 + \ub) z^6)}{9u^2(1 + \ub)z^2} \ln^2(z)
\no\\
&\quad
- \frac{2(72u^3 - (469 + 3\pi^2 - (59 - 3\pi^2)\ub)u z^2 + 12(29 + 3\ub
  + 14\ub^2)z^4)}{27u(1 + \ub)z^2} \ln(z)
-\frac49 \zeta_3
\no\\
&\quad
- \frac{( 12(7 - \ub)u^3 - 144u^3 z + 40(3 - \ub + 12\ub^2) z^4 - 36(1 +
  \ub)(1 - 8\ub + 3\ub^2) z^6)\pi^2}{54u^2(1 +\ub)z^2} 
\no\\
&\quad
- \frac{( 6(69 - 80\ub + 27\ub^2)u z^3- (163 - 189\ub)u^2 z^2 - 6(49 +
  119\ub + 175\ub^2 - 55\ub^3) z^5  )\pi^2}{54u^2(1 +\ub)z^2} 
\no\\
&\quad
- \frac{108(27 + 11\ub)u^3 - (5827 - 509\ub)u^2 z^2 +36(107 + 265\ub +
  395\ub^2 - 83\ub^3) z^4}{162 u^2 (1 + \ub) z^2} 
\no\\
&\quad
+ \frac{8((3 - 19\ub)u^2 + 2(15 - 23\ub)\ub z^2)}{9u^2\ub} \ln(z)\ln(u),
\no\\
& k_{12}(u, z) =
\frac{16z^2}{u^2} \calI_1(\ub, z)
+ \frac{8(u^2 + 4\ub z^2)}{3\ub(1 + \ub)} \calI_2(\ub, z)
- \frac{8( u^2 - 2 \ub z^2)}{3u\ub^2} \bigg( \Li_2(\ub) + \ln^2 (u) \bigg)
\no\\
&\quad
+\frac{8(u^3-2u^2z+(1 + 10\ub + \ub^2) z^2-4(5 + 2 \ub - \ub^2) \ub
  z^3+3(1+\ub)(3 - \ub) \ub z^4)}{3u^2 \ub(1 + \ub)} 
\no\\
&\quad
\times\bigg( \Li_2(z^2) + 2 \ln (z) \ln(1 - z^2) \bigg)
- \frac{16((3 - \ub) u + (13 + 5\ub)\ub z^2)}{3u \ub(1 + \ub)} \ln(z)
\no\\
&\quad
+ \frac{64(u^2z+2(5 + 2 \ub - \ub^2)\ub z^3)}{3u^2\ub(1 + \ub)}
\bigg( \Li_2(-z) + \ln (z) \ln(1 + z) \bigg)
\no\\
&\quad
- \frac{16(2 u^3 + (1 + 9\ub)u z^2 + 3(1+\ub)(3 - \ub) \ub
  z^4)}{3u^2\ub(1+\ub)} \ln^2 (z) 
- \frac{4(u^2+72 \ub z^2)}{9u\ub^2}  \ln (u)
\no\\
&\quad
- \frac{8( 2u^3 - 6 u^2 z + (1 + 17\ub) z^2 - 12 (5 + 2 \ub - \ub^2)\ub
  z^3 + 3(1+\ub) (3 - \ub) \ub z^4)\pi^2}{9u^2\ub(1+\ub)} 
\no\\
&\quad
- \frac{16(u^2 - 2\ub z^2)}{3u \ub^2} \ln(z) \ln(u)
- \frac{4((43 - 5\ub)u^2+6(86 + 41\ub -13\ub^2) \ub z^2)}{9u^2\ub(1+\ub)},
\no\\
& k_{13}(u, z) =
\frac{16(3u^3 + (37 + 15\ub + 2\ub^2)\ub z^2)}{3u^2\ub(1 + \ub)} \ln(z)
- \frac{16(1 + 2\ub) z^2}{u^3} \calI_1(\ub, z)
\no\\
&\quad
- \frac{8(1 - 2\ub)(u^2 + 4\ub z^2)}{3u \ub(1 + \ub)} \calI_2(\ub, z)
- \frac{16((1 - 2u) u^2 + 2(1 + 10\ub)\ub z^2)}{3u^2\ub^2} \ln(z)\ln(u)
\no\\
&\quad
- \frac{32((2 - \ub + 3\ub^2)u^2 z + (35 + 25\ub + 13\ub^2 - \ub^3)\ub
  z^3)}{3u^3\ub(1 + \ub)} 
\no\\
&\quad
\times \bigg( \Li_2(-z) + \ln (z) \ln(1+z) \bigg)
+ \frac{8((1 - 2\ub)u^2 - 2(1 + 10\ub)\ub z^2)}{3u^2\ub^2}
\bigg( \Li_2(\ub) + \ln^2(u) \bigg)
\no\\
&\quad
+ \frac{16(2(1 - 2\ub)u^3 + (1 + 19\ub - 6\ub^2)u z^2 + 3(1 + \ub)(5 +
  \ub)\ub z^4)}{3u^3\ub(1 + \ub)} \ln^2(z) 
\no\\
&\quad
- \frac{8(1 - z)^2((1 - 2\ub)u^3 - (5 - \ub)u^2 \ub z + 3(1 + \ub)(5 +
  \ub)\ub z^2)}{3 u^3 \ub(1 + \ub)} 
\no\\
&\quad
\times \bigg( \Li_2(z^2) + 2\ln (z) \ln(1-z^2) \bigg)
+ \frac{4((43 - 29 \ub)u^3 +6(140 + 127\ub + 79\ub^2 - 4\ub^3)\ub
  z^2)}{9u^3\ub(1 + \ub)} 
\no\\
&\quad
+ \frac{8(2(1 - 2\ub) u^3 - 3(2 - \ub + 3\ub^2)u^2 z + (1 + 27\ub +
  2\ub^2 + 24\ub^3)z^2)\pi^2}{9u^3\ub(1 + \ub)} 
\no\\
&\quad
- \frac{8(3(35 + 25\ub + 13\ub^2 - \ub^3) z^3 - 3(1 + \ub)(5 + \ub)
  z^4)\pi^2}{9u^3(1 + \ub)}
\no\\
&\quad
+ \frac{4((1 - 14\ub)u^2 + 72(1 + 2\ub)\ub  z^2)}{9u^2\ub^2} \ln(u),
\end{align}
with the finite terms of the 4-topology MIs $\calI_{1,2}(u,z)$ that we
introduced in Appendix~\ref{app:MIs}.

\newpage
\section{Master Integrals}
\label{app:MIs}

In this appendix we present our results for the MIs that involve the
charm quark mass. The MIs are normalized according to 
\begin{align}
[dk] \equiv  \frac{\Gamma(1-\eps)}{i\pi^{d/2}}\; d^d k
\end{align}
with $d=4-2\eps$. We suppress the $+i\eps$-prescription of the particle
propagators, 
\begin{align}
\calP_1 &= (k - l)^2 - z^2 m_b^2,
&&\calP_3 = (p + k)^2 - m_b^2,
\no\\
\calP_2 &= l^2 - z^2 m_b^2,
&&\calP_4 = (\bar u q + k)^2,
\end{align}
and write $z=m_c/m_b$ and $\ub=1-u$.

\subsubsection*{3--topologies}

\begin{align}
&\hspace{-4mm}
\parbox[c]{2.5cm}{\psfig{file=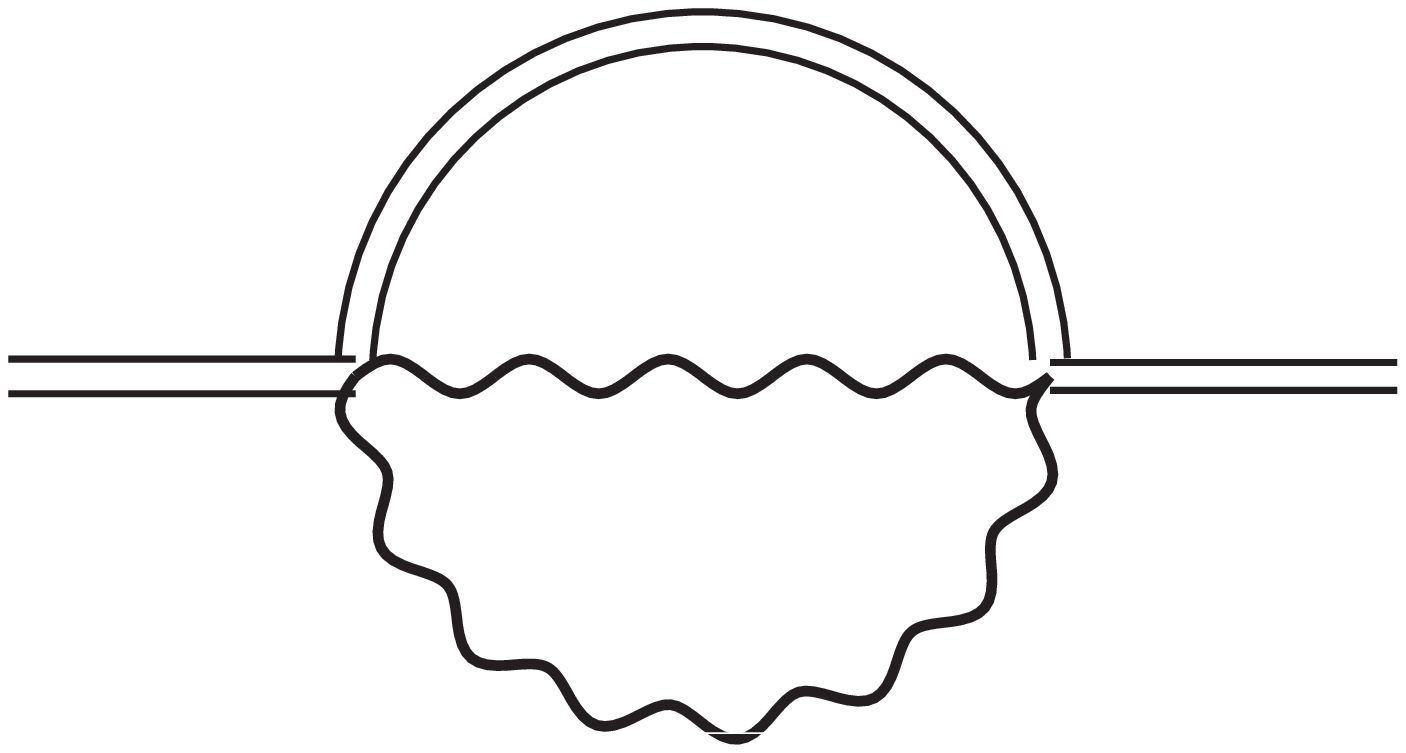,width = 2cm}}
=\quad
    \int [dk] [dl] \;  \frac{1}{\calP_{1}\calP_{2}\calP_{3}}
\quad \equiv \quad
    (m^2)^{1-2\eps} \; \left\{ \sum_{i=-2}^{1} d_i^{(31)} \, \eps^i +
      \calO(\eps^2) \right\} 
\end{align}
with
\begin{align}
d_{-2}^{(31)}   \, =&\;
    \frac12 + z^2,
    \no \\
d_{-1}^{(31)}    \, =&\;
    \frac54 + 3 z^2 - 4 z^2 \ln (z),
    \no \\
d_{0}^{(31)}    \, =&\;
    \frac{11}{8} + 6z^2
    + 2 z^2 \bigg( ( 2 + z^2 ) \ln(z)  - 7  \bigg)\ln(z)
     + \frac{(3 - 2 z^2 + 2 z^4)\pi^2}{6}
    \no \\
    &\;
    -(1-z^2)^2 \bigg( \text{Li}_2(z^2) + 2 \ln(z) \ln (1-z^2) \bigg),
    \no \\
d_{1}^{(31)}   \, =&\;
    - \frac{55}{16} + \frac{15}{2}z^2
    - \frac{z^2}{3} \bigg( (8 + 12z^2) \ln^2 (z) - 3(12 + 5z^2)\ln(z)
        + ( 111 + 4\pi^2) \bigg)\ln(z)
    \no \\
    &\;
    +2(1 - z^2)^2 \bigg[ 2\calH_1(z) + \S_{1,2}(z^2) + \ln(1 - z^2)
    \Li_2(z^2) + 2 \ln(1-z)\Li_2(-z) 
    \no \\
    &\;
    +\Big(2\ln^2(1-z^2) - \ln^2(1-z) - \ln^2(1 + z)\Big) \ln(z)
    - \pi^2 \ln(1 + z) + 2\zeta_3 \bigg]
    \no \\
    &\;
    +8z(1 + z^2)\bigg(\Li_2(-z) + \ln(z) \ln(1+z) \bigg)
    + \frac{( 15 + 24z - 24z^2 + 24z^3 + 10 z^4)\pi^2}{12}
    \no \\
    &\;
    - \frac12(1 - z)^2(5 + 14z + 5z^2)
        \bigg(\Li_2(z^2) + 2 \ln(z) \ln(1 - z^2) \bigg),
\end{align}
and the function $\calH_1(x)$ from Section~\ref{sec:2loop}.

\begin{align}
&\hspace{-4mm}
\parbox[c]{2.5cm}{\psfig{file=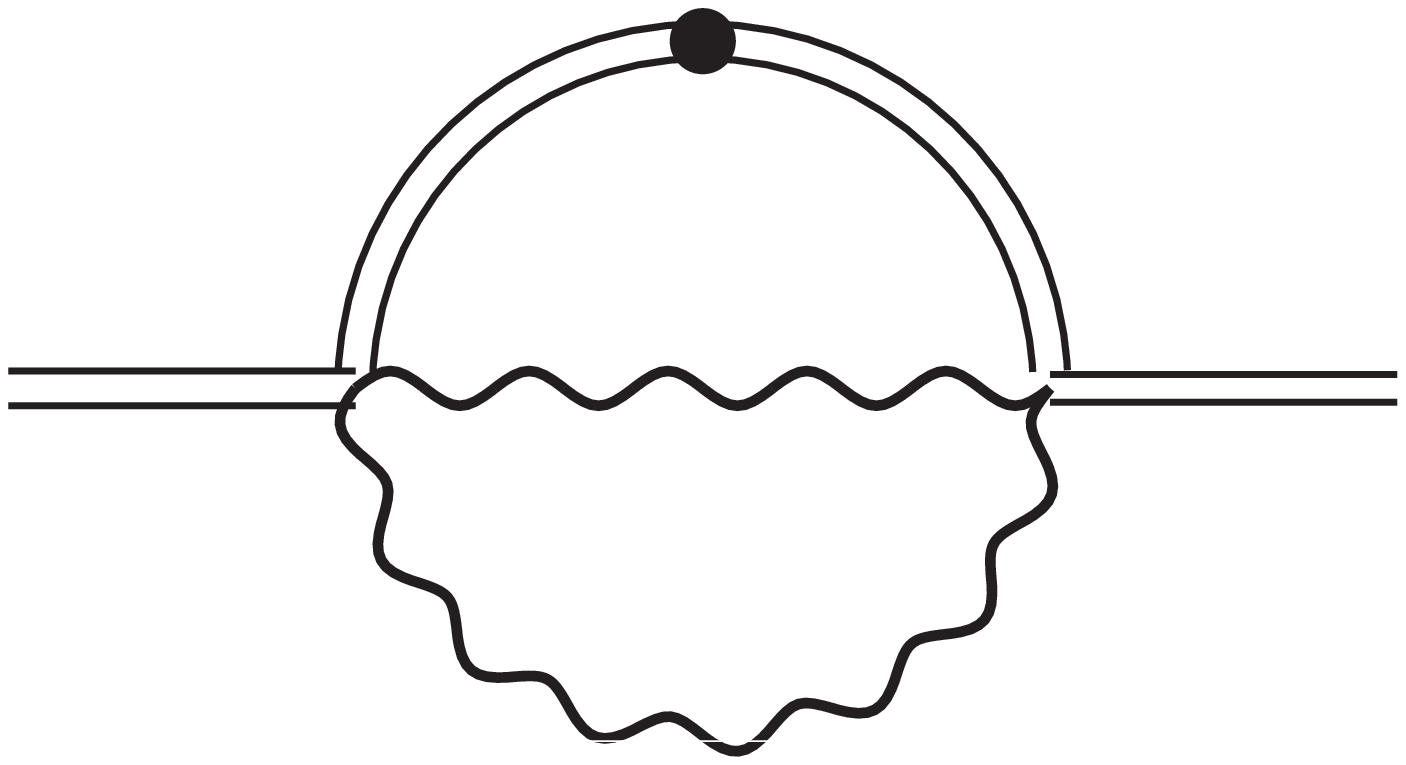,width = 2cm}}
=\quad
    \int [dk] [dl] \;  \frac{1}{\calP_{1}\calP_{2}\calP_{3}^2}
\quad \equiv \quad
    (m^2)^{-2\eps} \; \left\{ \sum_{i=-2}^{1} d_i^{(32)} \, \eps^i +
      \calO(\eps^2) \right\} 
\end{align}
with
\begin{align}
d_{-2}^{(32)}   \, =&\;
    \frac12, \no \\
d_{-1}^{(32)}    \, =&\;
    \frac12, \no \\
d_{0}^{(32)}    \, =&\;
    - \frac12 - 2 z^2 \ln^2 z
    -(1-z^2) \Bigg( \text{Li}_2(z^2) + 2 \ln(z) \ln (1-z^2) \bigg)
    + \frac{(3-2z^2)\pi^2}{6}, \no \\
d_{1}^{(32)}   \, =&\;
    -\frac{11}{2} + 2 z^2 \bigg( 2 \ln(z) - 3 \bigg) \ln^2 (z)
    + \frac{(1+4z-2z^2)\pi^2}{2}
    \no \\
    &\;
    +2(1 - z^2) \bigg[ 2\calH_1(z) + \S_{1,2}(z^2) + \ln(1 - z^2)
    \Li_2(z^2) + 2 \ln(1-z)\Li_2(-z) 
    \no \\
    &\;
    +8z\bigg(\Li_2(-z) + \ln(z) \ln(1+z) \bigg)\!
    - (1 - z)(1+3z)\bigg(\Li_2(z^2) + 2 \ln(z) \ln(1 - z^2) \bigg)
    \no \\
    &\;
    +\Big(2\ln^2(1-z^2) - \ln^2(1-z) - \ln^2(1 + z)\Big) \ln(z)
    - \pi^2 \ln(1 + z) + 2\zeta_3 \bigg].
\end{align}

\subsubsection*{4--topologies}

The 4-topology MIs are complicated since they depend on three scales
($um_b^2, m_c^2, m_b^2$). The divergent terms of the MIs are simple,  
\begin{align}
&\hspace{-4mm}
\parbox[c]{2.5cm}{\psfig{file=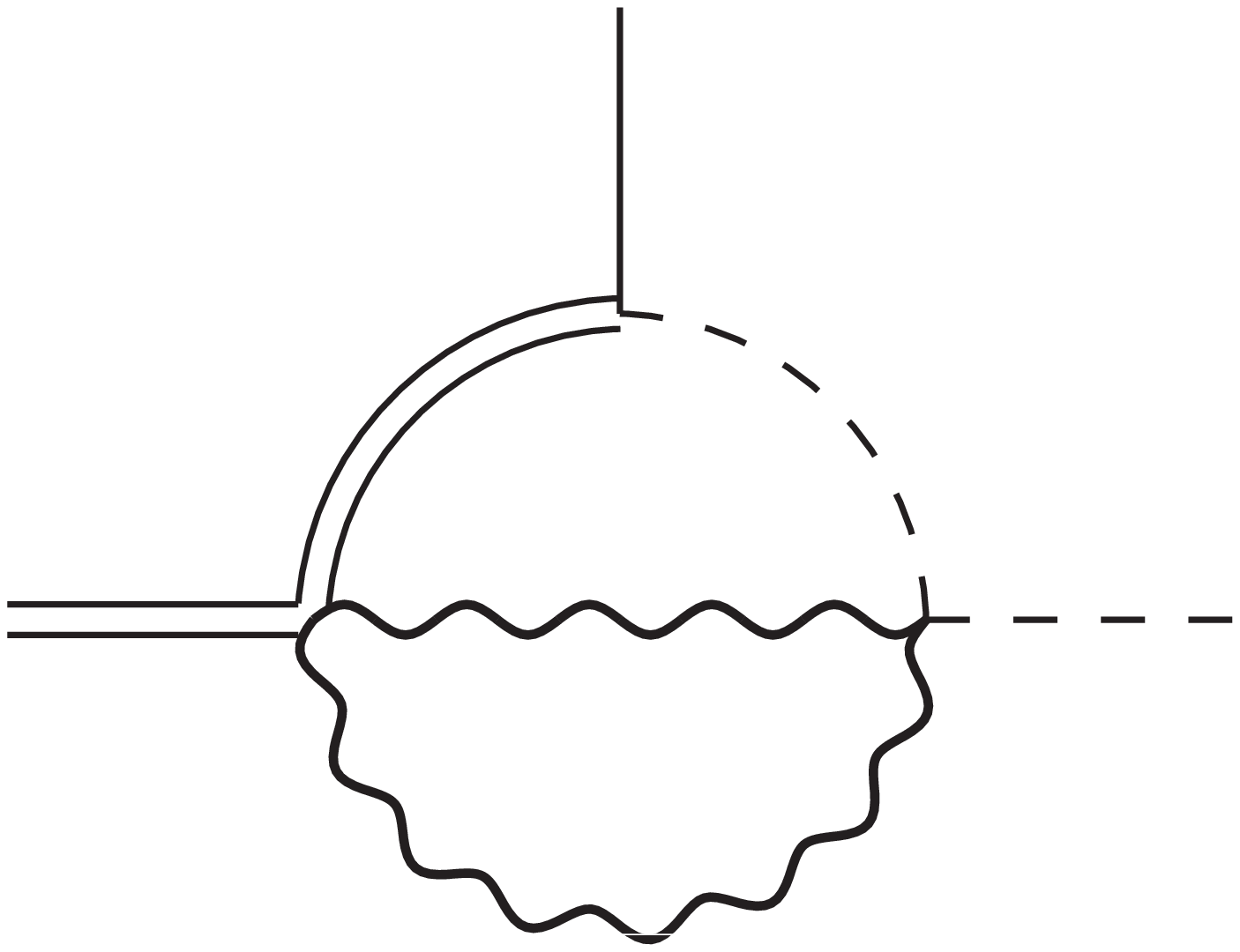,width = 2cm}}
=\quad
    \int [dk] [dl] \;  \frac{1}{\calP_{1}\calP_{2}\calP_{3}\calP_{4}}
\hspace{8.6cm}\no\\
&\hspace{2.1cm}= \quad
    (m^2)^{-2\eps} \; \left\{ \frac{1}{2\eps^2}
    +  \bigg(\frac{5}{2} + \frac{\ub}{u} \ln (\ub) \bigg)\frac{1}{\eps}+
    \calI_1(u,z) + \calO(\eps) \right\}, 
\end{align}
and
\begin{align}
&\hspace{-4mm}
\parbox[c]{2.5cm}{\psfig{file=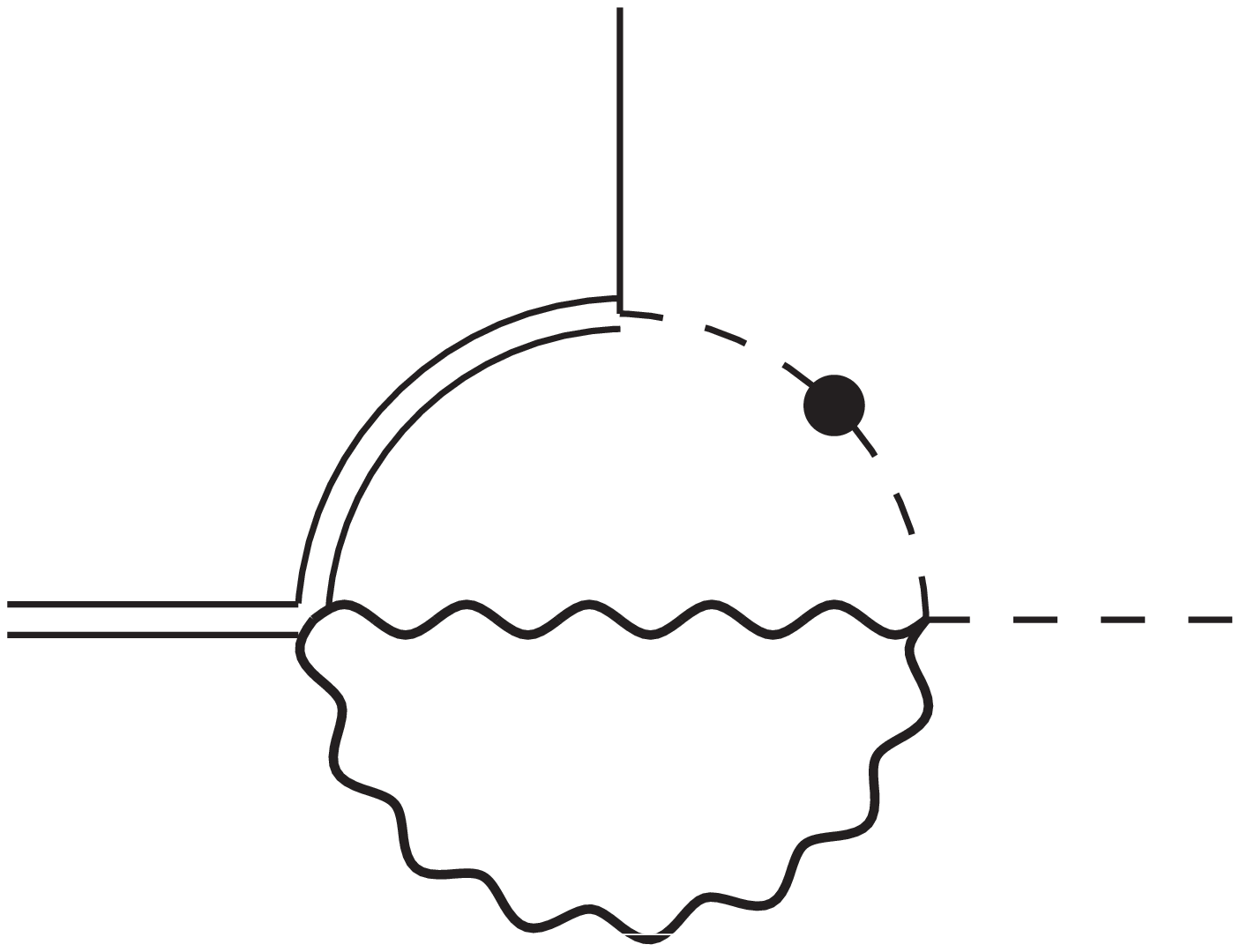,width = 2cm}}
=\quad
    \int [dk] [dl] \;  \frac{1}{\calP_{1}\calP_{2}\calP_{3}\calP_{4}^2}
\no\\
&\hspace{2.1cm}= \quad
    (m^2)^{-1-2\eps} \; \left\{ \frac{1}{\ub\eps^2}
    -  \bigg(2\ln(z) +  \frac{1+u}{u} \ln(\ub) \bigg)\frac{1}{\ub\eps}+
    \calI_2(u,z) + \calO(\eps) \right\}, 
\end{align}
but we could not find analytical expressions for the finite terms
$\calI_{1,2}(u,z)$\footnote{The structure of the differential equations
  indicates that $\calI_{1,2}(u,z)$ cannot be described by HPLs with
  simple arguments.}. Unfortunately, a power expansion in $z=m_c/m_b$
cannot be applied for the given MIs, since it generates powers of
$z/(1-u)$ which invalidate the expansion for large values of $u$. 

We could, however, derive suitable one-dimensional Mellin-Barnes
representations, which allow us to evaluate the finite terms numerically
to very high precision. We use this representation for our numerical
estimate of charm mass effects in Section~\ref{sec:numerics}, but also
provide compact parameterizations which are in any case convenient since
we expect the exact expressions to be rather involved. We propose the
parameterizations 
\begin{align}
\calI_{1}(u,z) \;&\simeq\; \frac{c_0 - c_1 u}{1 + (1 - u)^{\rho}},\no\\
\calI_{2}(u,z) \;&\simeq\; \frac{d_0}{(u_0-u)^\sigma(1-u)^\tau},
\label{eq:parametr}
\end{align}
which reproduce our numerical results over the entire $u$-range,
$0<u<1$,  and for physical values of $z=m_c/m_b$ at the percent
level. The numerical values of the parameters ($c_0,c_1,\rho,\ldots$)
depend on the value of $z$ and can be found in 
Table~\ref{tab:parameters}.

\begin{table}[h!]
\centerline{
\parbox{13cm}{\setlength{\doublerulesep}{0.1mm}
\centerline{\begin{tabular}{|c||c|c|c||c|c|c|c|} \hline
\hspace*{1cm}&\hspace*{1cm}&\hspace*{1cm}&\hspace*{1cm}&
\hspace*{1cm}&\hspace*{1cm}&\hspace*{1cm}&\hspace*{1cm}
\\[-0.7em]
$z$ & $c_0$ & $c_1$ & $\rho$
& $d_0$ & $u_0$ & $\sigma$
& $\tau$
\\[0.3em]
\hline\hline\hline&&&&&&&
\\[-0.7em]
$0.250$ &
$11.9022$ & $0.9512$ & $0.6575$ &
$13.1958$ & $1.0164$ & $0.2077$ &
$1.2244$
\\[0.3em]
\hline&&&&&&&
\\[-0.7em]
$0.275$ &
$11.5770$ & $0.9053$ & $0.6544$ &
$12.3101$ & $1.0171$ & $0.2170$ &
$1.2271$
\\[0.3em]
\hline&&&&&&&
\\[-0.7em]
$0.300$ &
$11.2538$ & $0.8523$ & $0.6506$ &
$11.5442$ & $1.0177$ & $0.2256$ &
$1.2295$
\\[0.3em]
\hline&&&&&&&
\\[-0.7em]
$0.325$  &
$10.9330$ & $0.7934$ & $0.6463$ &
$10.8767$ & $1.0182$ & $0.2335$ &
$1.2317$
\\[0.3em]
\hline&&&&&&&
\\[-0.7em]
$0.350$  &
$10.6147$ & $0.7293$ & $0.6414$ &
$10.2786$ & $1.0187$ & $0.2412$ &
$1.2339$
\\[0.3em]
\hline
\end{tabular}}
\vspace{4mm} \caption{\label{tab:parameters}\small \textit{Numerical
    input for the parameterizations (\ref{eq:parametr}) of the finite
    terms of the 4-topology Master Integrals.}}}} 
\end{table}

\end{appendix}



\begin{thebibliography}{99}

\bibitem{Chay:1990da}
  J.~Chay, H.~Georgi and B.~Grinstein,
  Phys.\ Lett.\  B {\bf 247} (1990) 399;\\
  I.~I.~Y.~Bigi, N.~G.~Uraltsev and A.~I.~Vainshtein,
  Phys.\ Lett.\  B {\bf 293} (1992) 430
  [Erratum-ibid.\  B {\bf 297} (1993) 477]
  [arXiv:hep-ph/9207214];\\
  I.~I.~Y.~Bigi, M.~A.~Shifman, N.~G.~Uraltsev and A.~I.~Vainshtein,
  Phys.\ Rev.\ Lett.\  {\bf 71} (1993) 496
  [arXiv:hep-ph/9304225];\\
  A.~V.~Manohar and M.~B.~Wise,
  Phys.\ Rev.\  D {\bf 49} (1994) 1310
  [arXiv:hep-ph/9308246].


\bibitem{BBNS}
  M.~Beneke, G.~Buchalla, M.~Neubert and C.~T.~Sachrajda,
  Phys.\ Rev.\ Lett.\  {\bf 83} (1999) 1914
  [arXiv:hep-ph/9905312];\\
  M.~Beneke, G.~Buchalla, M.~Neubert and C.~T.~Sachrajda,
  Nucl.\ Phys.\ B {\bf 591} (2000) 313
  [arXiv:hep-ph/0006124];\\
  M.~Beneke, G.~Buchalla, M.~Neubert and C.~T.~Sachrajda,
  Nucl.\ Phys.\ B {\bf 606} (2001) 245
  [arXiv:hep-ph/0104110].


\bibitem{SCET}
  C.~W.~Bauer, S.~Fleming, D.~Pirjol and I.~W.~Stewart,
  Phys.\ Rev.\  D {\bf 63} (2001) 114020
  [arXiv:hep-ph/0011336];\\
  C.~W.~Bauer, D.~Pirjol and I.~W.~Stewart,
  Phys.\ Rev.\  D {\bf 65} (2002) 054022
  [arXiv:hep-ph/0109045];\\
  M.~Beneke, A.~P.~Chapovsky, M.~Diehl and T.~Feldmann,
  Nucl.\ Phys.\  B {\bf 643} (2002) 431
  [arXiv:hep-ph/0206152].


\bibitem{Korchemsky:1994jb}
  G.~P.~Korchemsky and G.~Sterman,
  Phys.\ Lett.\  B {\bf 340} (1994) 96
  [arXiv:hep-ph/9407344].


\bibitem{ShapeFunc}
  M.~Neubert,
  Phys.\ Rev.\  D {\bf 49} (1994) 3392
  [arXiv:hep-ph/9311325];\\
  I.~I.~Y.~Bigi, M.~A.~Shifman, N.~G.~Uraltsev and A.~I.~Vainshtein,
  Int.\ J.\ Mod.\ Phys.\  A {\bf 9} (1994) 2467
  [arXiv:hep-ph/9312359].


\bibitem{NLO}
  C.~W.~Bauer and A.~V.~Manohar,
  Phys.\ Rev.\  D {\bf 70} (2004) 034024
  [arXiv:hep-ph/0312109];\\
   S.~W.~Bosch, B.~O.~Lange, M.~Neubert and G.~Paz,
   Nucl.\ Phys.\  B {\bf 699} (2004) 335
   [arXiv:hep-ph/0402094].


\bibitem{Becher:2006qw}
   T.~Becher and M.~Neubert,
   Phys.\ Lett.\  B {\bf 637} (2006) 251
   [arXiv:hep-ph/0603140].


\bibitem{Bonciani:2008wf}
  R.~Bonciani and A.~Ferroglia,
  JHEP {\bf 0811} (2008) 065
  [arXiv:0809.4687 [hep-ph]].


\bibitem{GB}
  G.~Bell,
  Nucl.\ Phys.\  B {\bf 795} (2008) 1
  [arXiv:0705.3127 [hep-ph]];\\
  G.~Bell, PhD thesis, LMU M\"unchen, 2006,
  arXiv:0705.3133 [hep-ph];\\
  G.~Bell,
  arXiv:0902.1915 [hep-ph].


\bibitem{IBP}
  F.~V.~Tkachov,
  Phys.\ Lett.\ B {\bf 100} (1981) 65;\\
  K.~G.~Chetyrkin and F.~V.~Tkachov,
  Nucl.\ Phys.\ B {\bf 192} (1981) 159.


\bibitem{Laporta}
  S.~Laporta,
  Int.\ J.\ Mod.\ Phys.\ A {\bf 15} (2000) 5087
  [arXiv:hep-ph/0102033].


\bibitem{DiffEqs}
  A.~V.~Kotikov,
  Phys.\ Lett.\ B {\bf 254} (1991) 158;\\
  E.~Remiddi,
  Nuovo Cim.\ A {\bf 110} (1997) 1435
  [arXiv:hep-th/9711188].


\bibitem{HPLs}
  E.~Remiddi and J.~A.~M.~Vermaseren,
  Int.\ J.\ Mod.\ Phys.\ A {\bf 15} (2000) 725
  [arXiv:hep-ph/9905237];\\
  D.~Maitre,
  Comput.\ Phys.\ Commun.\  {\bf 174} (2006) 222
  [arXiv:hep-ph/0507152].


\bibitem{MB}
  V.~A.~Smirnov,
  Phys.\ Lett.\ B {\bf 460} (1999) 397
  [arXiv:hep-ph/9905323];\\
  J.~B.~Tausk,
  Phys.\ Lett.\ B {\bf 469} (1999) 225
  [arXiv:hep-ph/9909506];\\
  M.~Czakon,
  Comput.\ Phys.\ Commun.\  {\bf 175} (2006) 559
  [arXiv:hep-ph/0511200].


\bibitem{SecDecomp}
  T.~Binoth and G.~Heinrich,
  Nucl.\ Phys.\ B {\bf 585} (2000) 741
  [arXiv:hep-ph/0004013].


\bibitem{FIESTA}
  A.~V.~Smirnov and M.~N.~Tentyukov,
  arXiv:0807.4129 [hep-ph].


\bibitem{Z2bOS:2loop}
  N.~Gray, D.~J.~Broadhurst, W.~Grafe and K.~Schilcher,
  Z.\ Phys.\  C {\bf 48} (1990) 673;\\
  D.~J.~Broadhurst, N.~Gray and K.~Schilcher,
  Z.\ Phys.\  C {\bf 52} (1991) 111.


\bibitem{Becher:2005pd}
  T.~Becher and M.~Neubert,
  Phys.\ Lett.\  B {\bf 633} (2006) 739
  [arXiv:hep-ph/0512208].


\bibitem{Chetyrkin:1997un}
  K.~G.~Chetyrkin, B.~A.~Kniehl and M.~Steinhauser,
  Nucl.\ Phys.\  B {\bf 510} (1998) 61
  [arXiv:hep-ph/9708255].


\bibitem{Korchemsky:1987wg}
  G.~P.~Korchemsky and A.~V.~Radyushkin,
  Nucl.\ Phys.\  B {\bf 283} (1987) 342;\\
  I.~A.~Korchemskaya and G.~P.~Korchemsky,
  Phys.\ Lett.\  B {\bf 287} (1992) 169.


\bibitem{Neubert:2004dd}
  M.~Neubert,
  Eur.\ Phys.\ J.\  C {\bf 40} (2005) 165
  [arXiv:hep-ph/0408179].


\bibitem{Asatrian:2008uk}
  H.~M.~Asatrian, C.~Greub and B.~D.~Pecjak,
  Phys.\ Rev.\  D {\bf 78} (2008) 114028
  [arXiv:0810.0987 [hep-ph]].


\bibitem{Beneke:2008ei}
  M.~Beneke, T.~Huber and X.~Q.~Li,
  Nucl.\ Phys.\  B {\bf 811} (2009) 77
  [arXiv:0810.1230 [hep-ph]].


\end{thebibliography}
\end{document}